\documentclass[iop, apj]{emulateapj}
\usepackage{amsmath,amssymb,amstext,float}
\usepackage[citecolor=black,linkcolor=black]{hyperref}
\usepackage{color}

\usepackage[all]{hypcap} %Links go to figures; breaks on deluxetables (use \capstartfalse \capstarttrue to fix it)
\usepackage{aas_macros}
\usepackage{natbib}
\shorttitle{Dynamics of Coronal-Hole Boundaries}
\shortauthors{Higginson et al.}

\newcommand{\kms}{{\rm km~s}$^{-1}$}
\newcommand{\hrs}{{\rm h}}

\newcommand\newgrl{{GeoRL}}%  % Geophysics Research Letters
\newcommand\newjcoph{{JCoPh}}%  % Journal of Computational Physics
\newcommand\newjgra{{JGRA}}%  % Journal of Geophysical Research A
\newcommand\newlrsp{{LRSP}}%  % Living Reviews in Solar Physics
\newcommand\newnat{{Natur}}%  % Nature
\newcommand\newprl{{PhRvL}}%  % Physical Review Letters
\newcommand\newphpl{{PhPl}}%  % Physics of Plasmas
\newcommand\newrmp{{RvMP}}%  % Reviews of Modern Physics
\newcommand\newsci{{Sci}}%  % Science
\newcommand\newsoph{{SoPh}}%  % Solar Physics
\newcommand\newssr{{SSRv}}%  % Space Science Reviews

\begin{document}

%%%%%%%%%%%%%%%%%%%%%%%%%%%%%%%%%%%%%%%%%%%%%%%%%%%%%%%%%%%%%%%%%%%%%%%%%%%%
\title{Dynamics of Coronal-Hole Boundaries}
%%%%%%%%%%%%%%%%%%%%%%%%%%%%%%%%%%%%%%%%%%%%%%%%%%%%%%%%%%%%%%%%%%%%%%%%%%%%

\author{A.~K.\ Higginson\altaffilmark{1}, S.~K.\ Antiochos\altaffilmark{2}, C.~R.\ DeVore\altaffilmark{2}, P.~F.\ Wyper\altaffilmark{3}, and T.~H.\ Zurbuchen\altaffilmark{1}}

\altaffiltext{1}{Department of Climate and Space Sciences and Engineering, University of Michigan, Ann Arbor, MI 48109, USA}

\altaffiltext{2}{Heliophysics Science Division, NASA Goddard Space Flight Center, Greenbelt, MD 20771, USA}

\altaffiltext{3}{Universities Space Research Association, NASA Goddard Space Flight Center, 8800 Greenbelt Road, Greenbelt, MD 20771, USA}

%%%%%%%%%%%%%%%%%%%%%%%%%%%%%%%%%%%%%%%%%%%%%%%%%%%%%%%%%%%%%%%%%%%%%%%%%%%%
\begin{abstract}
%%%%%%%%%%%%%%%%%%%%%%%%%%%%%%%%%%%%%%%%%%%%%%%%%%%%%%%%%%%%%%%%%%%%%%%%%%%%

Remote and in-situ observations strongly imply that the slow solar wind consists of plasma from the hot, closed-field corona that is released onto open magnetic field lines. The Separatrix Web (S-Web) theory for the slow wind proposes that photospheric motions, at the scale of supergranules, are responsible for generating dynamics at coronal-hole boundaries, which result in the closed plasma release. We use three-dimensional magnetohydrodynamic (3D MHD) simulations to determine the effect of photospheric flows on the open and closed magnetic flux of a model corona with a dipole magnetic field and an isothermal solar wind. A rotational surface motion is used to approximate photospheric supergranular driving and is applied at the boundary between the coronal hole and helmet streamer. The resulting dynamics consist primarily of prolific and efficient interchange reconnection between open and closed flux. Magnetic flux near the coronal-hole boundary experiences multiple interchange events, with some flux interchanging over fifty times in one day. Additionally, we find that the interchange reconnection occurs all along the coronal-hole boundary, even producing a lasting change in magnetic-field connectivity in regions that were not driven by the applied motions. Our results show that these dynamics should be ubiquitous in the Sun and heliosphere. We discuss the implications of our simulations for understanding the observed properties of the slow solar wind, with particular focus on the global-scale consequences of interchange reconnection.

\end{abstract}


\maketitle

%%%%%%%%%%%%%%%%%%%%%%%%%%%%%%%%%%%%%%%%%%%%%%%%%%%%%%%%%%%%%%%%%%%%%%%%%%%%
\section{Introduction}
%%%%%%%%%%%%%%%%%%%%%%%%%%%%%%%%%%%%%%%%%%%%%%%%%%%%%%%%%%%%%%%%%%%%%%%%%%%%

Understanding how the plasma and magnetic field of the Sun's
atmosphere -- from the photosphere to the coronas -- extend outward
to form the heliosphere has long been one of the central goals of
Heliophysics. In his pioneering work, \citet{parker58} gave the simplest
model for this Sun-heliosphere connection. Parker proved that for a
spherically symmetric atmosphere maintained at a roughly constant
temperature by some coronal heating process, the plasma would
expand outward to form a steady supersonic solar wind. Remote-sensing
observations of the solar corona, however, such as the exquisite
eclipse photographs of \citet{druckmuller09} and the ultra-high
resolution XUV images from \textit{Hinode} and the \textit{Solar Dynamics 
Observatory} \citep[e.g.][]{schrijver13}, show that the solar corona 
is very far from spherical symmetry due to the action of the Sun's 
magnetic field. The field adds both structure and dynamics to the 
basic picture proposed by Parker, affecting both its large-scale 
organization and also the thermal properties of the solar wind 
through small-scale dynamics.

The most fundamental structure introduced by the field is that it divides 
the corona into magnetically ``closed'' and ``open'' regions. In closed 
regions, the field lines have finite length and connect at two points 
down to the photosphere, confining the plasma.  In open regions, the 
field lines do not connect back to the photosphere within the inner corona and stretch 
out indefinitely to form the heliosphere. Note that, for a truly steady
state, the solar wind can originate only from the open-field regions.

These two types of coronal regions are readily apparent in X-ray/XUV
images of the Sun, because of the differences in the properties of
their plasma. In closed regions where the plasma is confined, both 
the density and temperature tend to be high, $N \sim 1 \times 10^9$
cm$^{-3}$ and $T \sim 1$ MK \citep{feldman78,laming97,warren09}, whereas in open regions, 
$N \sim 2 \times 10^8$ cm$^{-3}$ and $T \sim 0.8$ MK \citep{delzanna99,doschek97,landi08}. As a 
result, the open regions are often observed to be dark in X-ray images
and are referred to as ``coronal holes'' \citep{zirker77}.

Due to this temperature difference, the plasmas in open and closed
regions have different ionic charge-state composition, which is
readily seen in spectroscopic observations of the corona
\citep{doschek77,doschek97}. In addition, open and closed plasmas are observed
to have very different elemental composition. A key result emphasized
by \citet{meyer85} in his pioneering studies is that the elemental
abundances of coronal and heliospheric plasma are sensitive to the
first ionization potential (FIP) of the particular element. Many
studies have now shown conclusively that in open-field regions the
coronal plasma has heavy-element abundances close to those of the
photosphere. In closed regions, however, the low FIP elements such as
Fe and Mg have an abundance relative to the photosphere that is 4--6
times larger than the abundances of high FIP elements, such as N and
Ar \citep{meyer85,feldman03,laming15,vonsteiger16}.

This striking abundance variation also shows up in the plasma of the
solar wind.  It has long been known that the solar wind consists of
two distinct types, the so-called fast wind with speeds $> 500$ \kms
and the slow wind with speeds $< 500$ \kms. The fast wind is generally
believed to originate in coronal holes. For example, the Ulysses
measurements show that during solar minimum the wind at high latitudes
that emanates from the polar coronal holes is all fast wind 
\citep[][and references therein; \citeauthor{mccomas08} 
\citeyear{mccomas08}]{zurbuchen07}. 
Furthermore, this wind exhibits both elemental abundances
and ionic charge states indicative of a source back near the Sun that
is similar to coronal-hole plasma \citep{geiss95, vonsteiger00}. 
Latitudinal scans by Ulysses have shown that the 
fast wind exhibits little spatial variation in speed and composition 
over large coronal holes. 
Consequently, it can be thought of as the quasi-steady wind predicted by 
\citet{parker58}, although with different physical processes as its source.

The slow wind, on the other hand, has very different properties and
its source at the Sun is still an issue of intense debate.  It 
appears to be associated with streamers \citep{raymond97}, being limited to low 
latitudes during the minimum of solar cycle 23 \citep{manoharan12}, when the 
streamers were also found only at low latitudes. The slow wind extended further 
during the more complex solar minimum of cycle 24 \citep{tokumaru10}, when 
streamer structures also extended to higher latitudes. It is also associated 
with the heliospheric current sheet emanating from the top of the closed-field 
region \citep{gosling97, zhao09}. Indeed, the HCS is always 
embedded in slow wind, never fast \citep{burlaga02}. The location and 
composition of the slow wind suggest
that it is somehow associated with the closed-field regions.
Furthermore, the slow wind has elemental composition similar to that
of closed-field plasma, and its ionic charge states indicate a source
plasma with temperature similar to that of the closed corona rather
than coronal holes \citep{vonsteiger97, vonsteiger01, zurbuchen99, zurbuchen02}. 
In fact, several authors now argue that plasma composition is a much better 
discriminator between the two types of wind than the flow speed \citep[e.g.][]{zhao09}.

The third major feature of the slow wind that distinguishes it from
the fast is its variability. The slow wind exhibits continuous, strong
variability in all plasma properties, especially composition and
density \citep{gosling97}. This variability appears to be related to
specific structures and not the consequences of simple Alfv\'enic 
turbulence observed in the fast
wind. In fact, recent results by \citet{viall15} and \citet{kepko16}
indicate that 85\% or more of the slow wind consists of quasi-periodic
structures that vary rapidly, on time scales of hours, in both density
and plasma composition. Accompanying these plasma variations is strong
variability in the magnetic field measured at 1 AU with clear
signatures of plasmoids and disconnected flux \citep{kepko16}.
These authors conclude through a detailed analysis 
of white-light observations combined with in-situ measurements that 
magnetic reconnection in the streamer
stalks that map down to the closed-field region is the fundamental
process giving rise to the quasi-periodic structures.

As a result of the observations described above, especially the
composition and variability measurements, many models postulate that
slow wind is due to closed-field plasma that is released onto open
field lines \citep{antiochos11,fisk03}. This release is believed
to occur as a result of the magnetic-field dynamics, specifically
reconnection between open and closed flux -- so-called interchange
reconnection \citep{fisk99,crooker02}. There are
two main types of magnetically driven dynamical models for the origin
of the slow wind. In the interchange model proposed by Fisk and
co-workers, open flux is postulated to diffuse throughout the 
closed regions, releasing closed-field plasma via interchange
reconnection \citep{fisk98,fisk03,fisk09}. It should be noted,
however, that the fundamental tenet of this model -- open flux can
diffuse deep into closed regions -- has yet to be demonstrated by
rigorous numerical simulations.

A more widely accepted model is the streamer-top model 
\citep{suess96,sheeley97,endeve04,rappazzo05,sheeley09} and 
its extension, the Separatrix-Web (S-Web) model, in which 
the dynamics are localized at the boundary between open 
and closed flux \citep{antiochos07,antiochos11}.  The basic
idea of the S-Web model is that the driving of the coronal
magnetic field by continual photospheric motions must broaden 
any open-closed boundary, which for a steady state is a separatrix 
surface of zero width, into a finite-width dynamic boundary
layer. Magnetic flux in this boundary layer constantly opens and
closes as a result of interchange reconnection, with closed flux 
opening and open flux closing. This boundary layer extends out 
into the heliosphere to form the streamer stalks and embedded HCS. 

It should be noted that the dynamical driving of the open-closed
boundary has long been proposed to explain a number of observed
phenomena. On global scales, the continual opening and closing of
coronal-hole boundaries has been invoked \citep[e.g.][]{wang96} as the
mechanism that accounts for the apparent rigid rotation of some coronal 
holes \citep{timothy75,zirker77}. On intermediate scales,
interchange reconnection at the separatrices that define the tops of
pseudostreamers \citep{wang07} is widely believed to be responsible for
the bright plasma sheets in the heliosphere \citep{hundhausen72} that
define these structures. Finally, on small scales, reconnection
at the open-closed boundary separating small closed-field bipoles 
from the open flux of polar coronal holes is generally accepted 
to be the mechanism underlying a broad range of activity including
spicules, plumes, and jets \citep[e.g.][]{pariat15}.

Therefore, calculating and understanding
the dynamics of the open-closed boundary driven by photospheric
motions is essential for understanding a host of observed solar
activity and is at the heart of the S-Web model for the slow
wind. Since the photosphere is undergoing turbulent convection, its
flows cover all spatial scales ranging from the solar radius down to
the viscous dissipation scale, but for S-Web dynamics we can consider
these flows to have three dominant scale regimes. For global motions
such as the differential rotation or meridional flow that have time
scales of many days, much longer than the typical time scale for
setting up a steady wind ($\sim$ one day), the open-closed
boundary evolution can be considered as quasi-steady. In the other
extreme, the small-scale flows of granules and the magnetic carpet
\citep{schrijver97}, which are of order minutes, are much smaller 
than a day, so these are likely to produce only Alfv\'en wave 
``noise'' at the boundary. This noise may play a critical role in
heating the plasma and driving the wind, but we do not expect it to be
important for releasing the closed-field plasma of the slow wind.  The
important motions for driving the open-closed boundary are likely to
be the supergranular flows, which have time scales of order one day
or so. 

In this paper, we present the first detailed calculations of the dynamical
response of a prototypical coronal-hole boundary to driving 
by supergranule-like motions. 
To understand the overall topology and physical mechanisms, these first 
calculations use the simplest structure for the initial coronal magnetic field
and the driving motions. As with all numerical experiments, we 
are limited to a Lundquist number that is orders of magnitude smaller than solar 
values. This and other numerical issues are addressed below in \S \ref{conclusions}. 
However, we have carefully designed our simulations so that our results can be 
applied with confidence to the Sun. Although we do not attempt to match exactly 
the slow speed and small scale of photospheric supergranules, we will show that, 
even with the simplest possible initial open-closed 
boundary and representation of photospheric driving, 
the system quickly develops intricate three-dimensional (3D) structure. 
This results in complex reconnection dynamics between open and closed flux 
that should be present at any boundary between a coronal hole 
and a helmet streamer, lending support to S-Web-type models. We also
emphasize that, irrespective of the theoretical model for the slow 
solar wind, the driving of coronal-hole boundaries by photospheric 
motions must be a generic feature of the Sun's corona, and must 
be considered when interpreting observations of the corona and wind. As discussed 
below, the results of this first investigation into these generic dynamics have 
important implications for understanding how the corona produces the solar wind 
that we observe.

%%%%%%%%%%%%%%%%%%%%%%%%%%%%%%%%%%%%%%%%%%%%%%%%%%%%%%%%%%%%%%%%%%%%%%%%%%%%
\section{Method}
%%%%%%%%%%%%%%%%%%%%%%%%%%%%%%%%%%%%%%%%%%%%%%%%%%%%%%%%%%%%%%%%%%%%%%%%%%%%
\subsection{MHD Model and Initial Conditions} \label{model}
%%%%%%%%%%%%%%%%%%%%%%%%%%%%%%%%%%%%%%%%%%%%%%%%%%%%%%%%%%%%%%%%%%%%%%%%%%%%

Our calculations use the Adaptively Refined Magnetohydrodynamic
(MHD) Solver (ARMS), which uses Flux Corrected Transport methods to
capture shocks and minimize diffusion \citep{devore91}. Our numerical
domain consists  of a spherical wedge centered on the Sun, in which we 
solve the following ideal MHD equations: 
%%%%%%%%%%%%%%
\begin{eqnarray}
%%%%%%%%%%%%%%
\label{continuityequation}
%%%%%%%%%%%%%%
\frac{\partial \rho}{\partial t} + {\bf \nabla} \cdot (\rho {\bf u}) &=&0,\\
%%%%%%%%%%%%%%
\label{momentumequation}
%%%%%%%%%%%%%%
\frac{\partial \rho {\bf u}}{\partial t} +
  {\bf \nabla} \cdot \left( \rho{\bf u} {\bf u} \right)&=& \frac{1}{4\pi}\left( {\bf \nabla} \times {\bf B} 
\right )\times {\bf B}  -  {\bf \nabla} P + \rho{\bf g} ,\\
%%%%%%%%%%%%%%
\label{inductionequation}
%%%%%%%%%%%%%%
 \frac{\partial {\bf B}}{\partial t} - {\bf \nabla} \times
\left(         {\bf u} \times {\bf B} \right) &=& 0,
%%%%%%%%%%%%%%
\end{eqnarray}
%%%%%%%%%%%%%%
\noindent 
where $\rho$ is the plasma mass density, $\bf u$ is the plasma velocity,
$\bf B$ is the magnetic field, $P$ is the plasma pressure, and ${\bf g} 
= -GM_\odot{\bf r}/r^3$ is the gravitational acceleration. The
plasma pressure is calculated from $P = 2 (\rho / m_p) k_B T$, i.e.\ we
assume the simplest solar atmosphere of a fully-ionized hydrogen gas
at constant, uniform temperature, $T = 1 $MK. ARMS stores the variables 
on a staggered grid to keep the divergence of the magnetic field fixed 
at zero to machine precision. It solves the equations using a second-order 
predictor/corrector in time and a fourth-order integrator in space. Together 
with the flux limiter, this ensures that magnetic reconnection via numerical 
diffusion takes place only in regions where the magnetic field develops 
structure down to the scale of the grid. In these locations, the effective 
numerical resistivity is determined by the flow speed and grid spacing. Since 
the main numerical challenge for our simulations is to resolve as much structure 
as possible with limited computational resources, we do not include an explicit 
resistivity.

The initial magnetic field is calculated using the Potential Field
Source Surface (PFSS) model \citep{altschuler69} for a dipole at Sun
center. The dipole strength is chosen so that the magnetic field at 1
$R_\odot$ is equal to 10 G at the poles, which is a good estimate
for the quiet-Sun magnetic field \citep{long13}. The radius of the
source surface, beyond which the initial magnetic field is assumed 
to be radial, is set at $R_S = 3 R_\odot$ \citep{schatten69}. 
As discussed below, this initial field is then 
allowed to equilibrate with the initial atmosphere and solar wind 
to reach a new quasi-steady equilibrium before the system experiences 
any driving.

For the initial atmosphere, we use Parker's isothermal, transonic
solar-wind solution \citep{parker58}. The velocity of the steady,
isothermal solar wind is given by 
%%%%%%%%%%%%%%
\begin{eqnarray}
%%%%%%%%%%%%%%
\label{eq-parker}
%%%%%%%%%%%%%%
\frac{v^2(r)}{c_s^2} \exp \left(1- \frac{v^2(r)}{c_s^2} \right) = 
\frac{r_s^4}{r^4} \exp \left( 4 - 4\frac{r_s}{r} \right),
%%%%%%%%%%%%%%
\end{eqnarray}
%%%%%%%%%%%%%%
\noindent where $v(r)$ is the solar wind velocity, $c_s^2 = 2 k_B T_0 / m_p$ 
is the sound speed, and $r_s = G M_{\odot} m_p / 4 k_B T_0$ is the sonic 
point. The plasma number density at the base of the atmosphere is a free 
parameter that we set to $3.6 \times 10^9$ cm$^{-3}$ to yield densities 
of $10^8$--$10^9$ cm$^{-3}$ in the helmet streamer. 

\begin{figure*}[!htpb]
%\centering\includegraphics[scale=0.4,trim=0cm 0cm 4.5cm 0cm, clip=true]
\centering\includegraphics[width=0.85\textwidth]{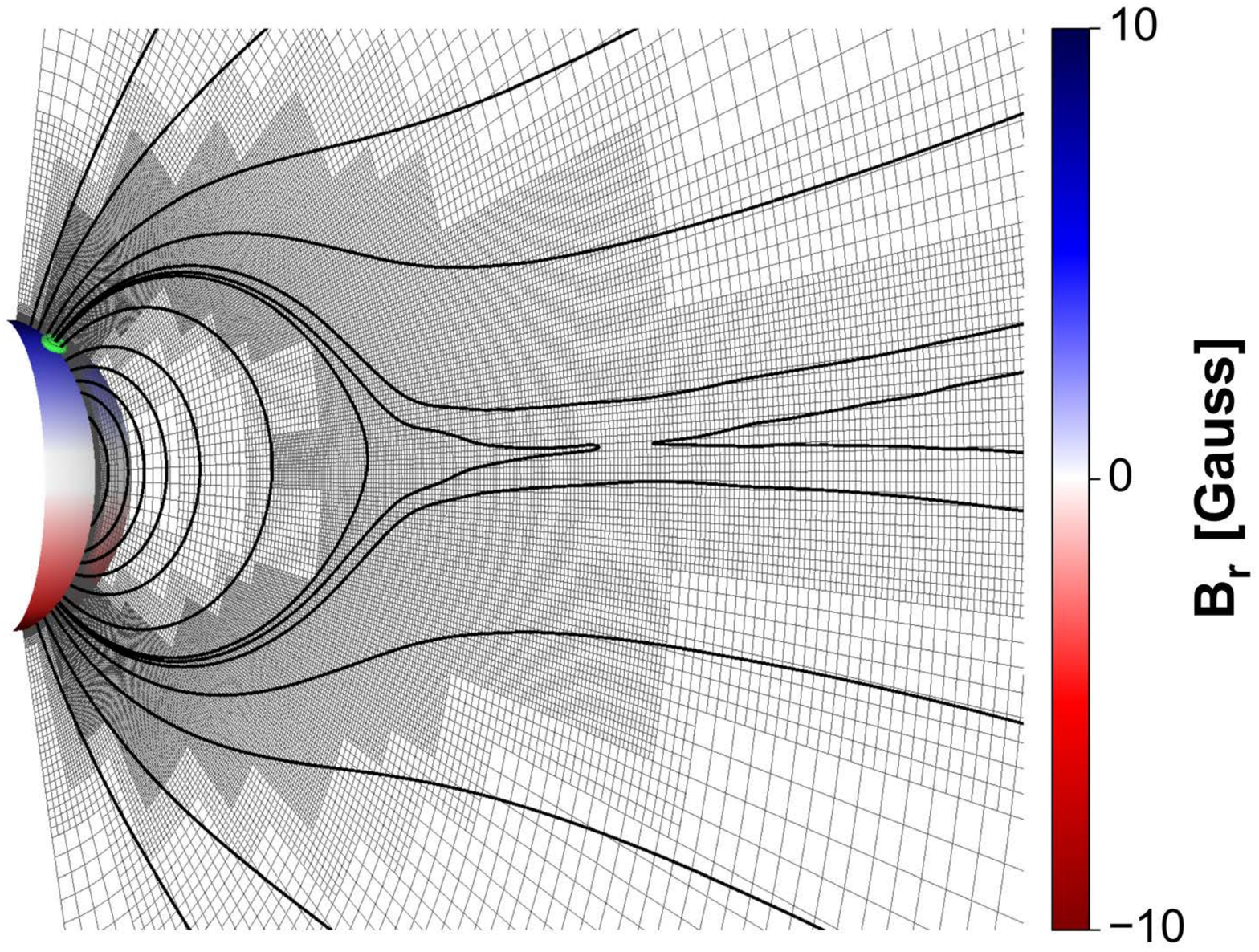}

\caption{Snapshot of the domain after dynamic relaxation to a quasi-steady state. 
Grid blocks (thin black lines) are shown in the plane $\phi=0$. Magnetic field lines 
(thick black curves) in this plane outline a dipolar magnetic field with two polar 
coronal holes and a helmet streamer that has recently pinched off at the top near 
the heliospheric current sheet. The solar surface is color-shaded according to the 
radial magnetic field component. Green contours in the northern hemisphere show the 
location of the driving-flow annulus straddling the coronal-hole boundary, and are 
drawn where the tangential velocity magnitude $\vert {\bf V} \vert = 4$ \kms.}

\label{figure1}

%%%%%%%%%%%%%%%%%%%%%%%%%%%%%%%%%%%%%%%%%%%%%%%%%%%%%%%%%%%%%%%%%%%%%%%%%%%%
\end{figure*}

Figure \ref{figure1} shows the spherical wedge domain, for which $R \in 
[1R_\odot, 30R_\odot]$, $\theta \in [11.25^{\circ}, 78.75^{\circ}]$, and 
$\phi \in [-22.5^{\circ}, +22.5^{\circ}]$. By setting our 
top boundary at $30R_\odot$, well above the top of the helmet streamer, we 
are able to simulate the entire radial extent of the coronal-hole boundary 
and its transition into the heliospheric current sheet. Shown on the radial 
surface in Figure \ref{figure1}
is a map of $B_r$, with green contours drawn to mark the location of 
the driving (see \S \ref{driver}). Pictured along a slice of constant
longitude are the block boundaries, where each block contains $8 \times 8 
\times 8$ grid cells. The grid is logarithmically stretched in $R$. We use 
an adaptive grid with the highest resolution along the entire coronal-hole 
boundary and the HCS, as well as a shallow layer of high-resolution cells
at the base to capture the velocity gradients where the solar wind is 
accelerated and the footpoint driving is imposed. The finest grid at 
the surface has $\sim$ 45 points per square degree. The coarsest grids 
are positioned in the corners of the wedge to save computational resources.

The radial inner boundary is set to be an effusing wall, which allows 
mass to flow into, but not out of, the system. Here we also line-tie 
the magnetic field by setting the tangential velocity everywhere on 
the surface to zero, except for the prescribed flow profile described 
below (\S \ref{driver}). The three radial guard cells beneath the lower 
boundary are held fixed at their initial densities, and their velocities 
are set to zero. All other boundaries allow mass to flow through, with the 
density and magnetic fields extrapolated using zero-gradient conditions. 
The velocities at the radial outer boundary assume free-flow-through (zero 
gradient) for the normal component and semi-slip conditions (zero value 
outside the boundary) on the tangential components. At the four side walls, 
the velocities are set to semi-flow-through (zero value outside) for the 
normal component and free-slip (zero gradient) on the tangential components.
These boundary conditions applied to the initial atmosphere produce a 
self-sustaining, isothermal solar wind throughout the domain, replenished 
from below in the open-field regions that have sustained upflows.

As mentioned above, the initial magnetic field (determined 
from the PFSS solution) and solar 
wind are not in equilibrium. Consequently, the first part of our simulation 
is a relaxation phase where the solar-wind plasma and magnetic field evolve 
toward a new pressure balance.  This relaxed condition is shown in Figure 
\ref{figure1}, where the thick black magnetic field lines outline the coronal 
holes at both poles, a stretched helmet streamer at the equator, and a dynamic 
heliospheric current sheet.

We find that this system never reaches a true time-independent steady state, 
because reconnection continually occurs in the HCS near the top of the helmet 
streamer. Due to the action of the solar wind, which stretches field lines out 
to infinity, the HCS continuously thins down until eventually its width reaches 
the grid scale. Reconnection at the HCS then forms a flux-rope-like structure 
that is carried outward with the solar wind. In our simulation, the reconnection 
is determined by the effective numerical resistivity (i.e., the Lundquist number), 
which scales directly with the grid spacing. Consequently, the non-steady 
dynamics become smaller and occur less frequently as the refinement 
increases. Scaling our results to solar Lundquist numbers, we expect that the 
non-steady background dynamics due to this HCS non-equilibrium 
reconnection process would have no observable consequences in imaging data. 
Thus, this process is not responsible for the formation of the 
quasi-periodic, slow-wind structures discovered by \citet{viall15} and \citet{kepko16}. 
In our simulation, the non-steady background dynamics merely constitute a very low 
level of fluctuations above which we easily detect the far larger effects of our 
surface driving. Moreover, that driving has a fixed scale determined by the size 
of observed supergranules and, hence, does not depend upon numerical refinement.

%-- \textcolor{red}{much like} a streamer blob \citep[e.g.][]{sheeley09} -- 
%%%%%%%%%%%%%%%%%%%%%%%%%%%%%%%%%%%%%%%%%%%%%%%%%%%%%%%%%%%%%%%%%%%%%%%%%%%%
\subsection{Boundary Driver} \label{driver}
%%%%%%%%%%%%%%%%%%%%%%%%%%%%%%%%%%%%%%%%%%%%%%%%%%%%%%%%%%%%%%%%%%%%%%%%%%%%

The  base  dynamics in  the  heliosphere  must  be determined  by  the
never-ceasing, supergranular-scale, photospheric convection. The actual
photospheric horizontal flows due to supergranules are highly complex,
as  convective cells appear and disappear randomly throughout the
photosphere.  Furthermore, the motions have both compressible and
incompressible components, with the flows expanding out radially from
cell centers and converging onto the network, where the misalignment 
of the flows and their random temporal dependence gives rise to
incompressible vortical motions with typical flow speeds of order 1 
\kms \citep{brandt88, duvall00, gizon03, komm07, attie09, seligman14}. 
For injecting stress into the coronal field, the  most important motions 
are the rotational components. Therefore, for this first investigation of 
supergranular driving, we use a simple model for a single supergranular flow. 

We impose a circular photospheric flow straddling the coronal-hole boundary 
(the location shown by the green contours on the radial surface in Figure 
\ref{figure1}). This flow lies in the $\theta$,$\phi$ plane only and is 
constructed so as to preserve the normal component of the magnetic field 
during the evolution. In order to satisfy 
%%%%%%%%%%%%%%
\begin{eqnarray}
%%%%%%%%%%%%%%
\label{FlowInductionEqn}
%%%%%%%%%%%%%%
\frac{\partial B_r}{\partial t} = - \nabla_\perp \cdot \left( \mathbf{v_{\perp}} B_r \right) = 0,
%%%%%%%%%%%%%%
\end{eqnarray}
%%%%%%%%%%%%%%
\noindent
we choose $\bf{v_\perp}$ to be equal to the curl of a radial vector, 
%%%%%%%%%%%%%%
\begin{eqnarray}
%%%%%%%%%%%%%%
\label{CurlVector}
%%%%%%%%%%%%%%
\bf{v_\perp} = \nabla_\perp \times \left( \psi, 0, 0 \right).
%%%%%%%%%%%%%%
\end{eqnarray}
%%%%%%%%%%%%%%
\noindent We define $\psi$ as a function of $\theta$, $\phi$, and $t$, 
%%%%%%%%%%%%%%
\begin{eqnarray}
%%%%%%%%%%%%%%
\label{psi}
%%%%%%%%%%%%%%
\psi \left( \theta, \phi, t \right) \equiv V_0 f(t) g(\xi) h(\beta),
%%%%%%%%%%%%%%
\end{eqnarray}
%%%%%%%%%%%%%%
\noindent
where 
%%%%%%%%%%%%%%
\begin{eqnarray}
%%%%%%%%%%%%%%
\label{ComponentsOfPsi}
%%%%%%%%%%%%%%
f(t) &=& \frac{1}{2} \left[ 1 - \cos \left( 2 \pi k \frac{t-t_0}{t_2 - t_1} \right) \right], \\
g(\xi) &=& \frac{(m + l + 1)}{(l+1)} \left[1-\xi^{2(l+1)} \right] - \left[ 1 - \xi^{2(m+l+1)}\right],
\\
h(\beta) &=& \frac{1}{2}\beta^2.
%%%%%%%%%%%%%%
\end{eqnarray}
%%%%%%%%%%%%%%

\noindent In the equations above, the parameters $k$, $t_0$, $t_1$, $t_2$ are set 
to ramp up the flow to maximum velocity from zero and then from that 
velocity back to zero. This ensures that all disturbances are smooth. 
The equation for $g(\xi)$ defines an annulus in spatial coordinate 
$\xi$, where

%%%%%%%%%%%%%%
\begin{eqnarray}
%%%%%%%%%%%%%%
\xi^{2} \equiv 4 \left( \frac{\theta - \theta_{0}}{\theta_{2} - \theta_{1}} \right) ^{2} + 4 
\left( \frac{\phi - \phi_{0}}{\phi_{2} - \phi_{1}} \right) ^{2}.
%%%%%%%%%%%%%%
\end{eqnarray}
%%%%%%%%%%%%%%

\noindent The location of the flow annulus is determined by the limits $\theta_1$, $\theta_2$, 
and $\phi_1$, $\phi_2$, with coordinate $\left( \theta_0,\phi_0 \right)$ representing the center. 
The thickness and radial velocity profile of the flow annulus are defined by $m$ 
and $l$. We set $m = l = 1$ to yield a thick annulus with a velocity 
peak at the center of the annulus.  In the equation for $h(\beta)$, 
$\beta$ is the magnetic field coordinate between minimum and maximum 
strengths, i.e., 

%%%%%%%%%%%%%%
\begin{eqnarray}
%%%%%%%%%%%%%%
\beta \equiv \max \left( \min \left( B_{r}, B_{2} \right) , B_{1}\right).
%%%%%%%%%%%%%%
\end{eqnarray}
%%%%%%%%%%%%%%
where we chose $B_{1} = 0$ G and $B_{2} = 10.0$ G so that $\beta = B_{r}$ everywhere in our flow region.  

\begin{figure}[!htbp]

\includegraphics[width=0.5\textwidth]{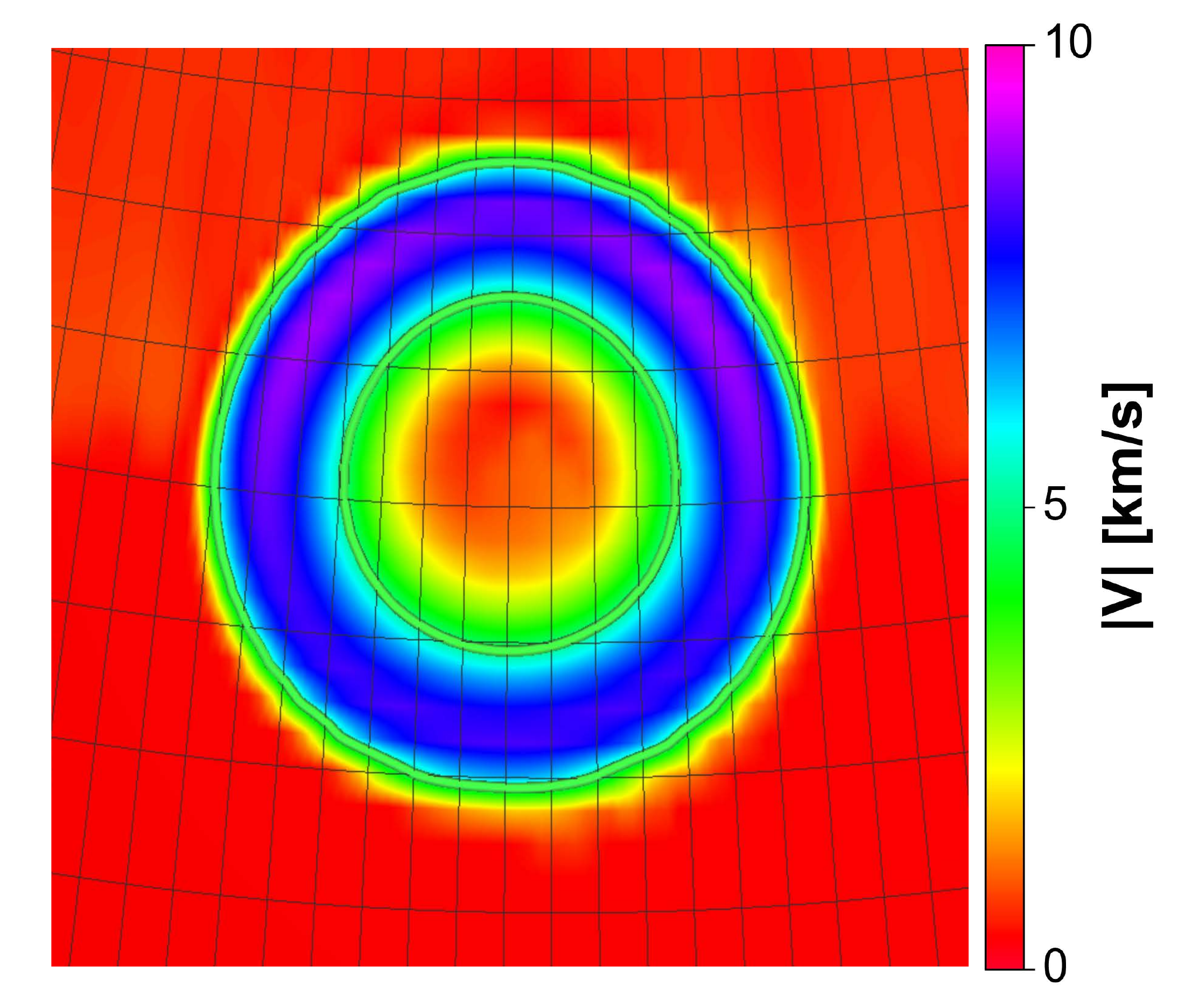}

\caption{Zoom-in on the bottom radial boundary ($R = 1 R_{\odot}$) showing tangential 
velocity magnitude at the peak of the driving in color shading. The green contours match 
the annuli shown in Figure \ref{figure1}. Block boundaries (thin black lines) also are 
shown.}

\label{figure2}
%%%%%%%%%%%%%%%%%%%%%%%%%%%%%%%%%%%%%%%%%%%%%%%%%%%%%%%%%%%%%%%%%%%%%%%%%%%%
\end{figure}

A contour map of tangential velocity on the surface is shown in Figure
\ref{figure2}, where the green velocity contours from Figure \ref{figure1} 
are shown for context. On the surface, $V_R \approx 0$ so that $|V|$ is 
essentially the tangential velocity only. The thin black lines show the 
block boundaries. The flow spans $40 \times 120$ grid points and has a 
diameter of about $1 \times 10^5$ km. By comparison, a typical supergranule 
cell has a diameter of $3 \times 10^4$ km \citep{simon64}. 

We adopt the 
larger size in order to highly resolve the flow on our grid. The cell is 
centered at $54.5^\circ$ above the equator on the coronal-hole boundary 
and at $0^\circ$ longitude. The flow extends from $50.4^\circ$ to $58.6^\circ$, 
for a total width of $8.2^\circ$ in latitude. In longitude, it extends from 
$-7^\circ$ to $+7^\circ$, for a total width of $14^\circ$ in longitude. 
The parameter $V_0$ in Equation \ref{psi} is chosen so that the maximum 
flow speed is 9.1 \kms. We ramp the flow up from zero to this maximum 
speed, drive it steadily, and then ramp it back down to zero. The maximum 
speed, 9.1 \kms, is only 3\% of the simulated Alfv\'en speed on the surface, 
which is of the same order of magnitude as for the Sun. By placing this
flow so that it straddles the coronal-hole boundary, we displace both 
open and closed magnetic flux from their original equilibrium positions, 
just as the supergranular flows must distort coronal-hole boundaries on 
the Sun.

To gain physical insight into the open-closed dynamics, we simulated two
cases that are presented below. On the Sun, both the southern and
northern coronal-hole boundaries are driven continuously by random,
out-of-phase motions. Our goal, however, is to understand in detail 
the fundamental dynamics of the boundary. Consequently, we designed 
the simulations so as to isolate the basic effect of the driving with
minimal extraneous complexity. We simulated a large 
polar coronal hole in order to maximize the effective numerical resolution 
available to us. However, our results apply to any coronal hole that is 
larger than the scale of the driver and flanks the helmet streamer, 
regardless of latitude. As will be evident below, even the simplest
driver results in highly complex dynamics. In both cases, we displaced 
the coronal-hole boundary only in the northern hemisphere, using the flow 
pattern described above.  In the first case, we twisted the field-line 
footpoints through a maximum angle of $\pi/2$, i.e.\ a quarter rotation. 
The ramp-up and ramp-down of the driving lasted for 2.4 \hrs.  Afterwards, 
all tangential velocities on the surface were again set to zero, and the 
system was allowed to relax. This relaxation phase extended to $t = 30.3$ 
\hrs, where $t = 0$ \hrs \;marks the start time of the driving. This finite, 
but not extreme, distortion of the boundary allows us to examine in detail 
how the system evolves. Then, in the second case, to better simulate the 
complexity resulting from constant photospheric driving, we twisted the 
footpoints through a maximum angle of $2\pi$, i.e.\ a full rotation. This 
motion results in a very complex distortion of the initial boundary, more 
like the actual supergranular flow is expected to produce. The driving phase 
lasts 6.0 \hrs, and the relaxation extends to $t = 75.6$ \hrs. Below, we 
present the results of these two simulations.

%%%%%%%%%%%%%%%%%%%%%%%%%%%%%%%%%%%%%%%%%%%%%%%%%%%%%%%%%%%%%%%%%%%%%%%%%%%%
\section{Results}
%%%%%%%%%%%%%%%%%%%%%%%%%%%%%%%%%%%%%%%%%%%%%%%%%%%%%%%%%%%%%%%%%%%%%%%%%%%%
\subsection{$\pi/2$ Displacement}
%%%%%%%%%%%%%%%%%%%%%%%%%%%%%%%%%%%%%%%%%%%%%%%%%%%%%%%%%%%%%%%%%%%%%%%%%%%%

Figures \ref{figure3} and \ref{figure4} show the imprint of 
the coronal-hole boundary on the solar surface for the $\pi/2$ 
rotation at various times during the evolution. 
Here we show snapshots before driving at $t = -0.03$ \hrs, immediately after 
driving at $t = 2.4$ \hrs, and then at three additional times chosen to best 
illustrate the phases of evolution, at times $t = 11.6$ \hrs, $t = 21.0$ \hrs,
and $t = 30.3$ \hrs. (We strongly encourage the reader to download the full, 
5-minute-cadence movies that are available online.) At each time, the coronal-hole 
boundary is shown in both the northern (left) and southern (right) hemispheres. 
Only the northern boundary is driven by the rotational flow. In these plots, 
white represents open magnetic field and black represents closed. The closed 
field in the north always connects to the closed field in the south.

To distinguish open and closed field, we traced a $1000 \times 1818$ grid of 
magnetic field lines distributed over $+35^\circ$ to $+75^\circ$ latitude and 
$-20^\circ$ to $+20^\circ$ longitude in the north, and $-75^\circ$ to $-35^\circ$ 
latitude and $-20^\circ$ to $+20^\circ$ longitude in the south. Open field is 
defined as those field lines that reach past $12 R_\odot$, where the solar-wind 
speed becomes greater than the Alfv\'en speed. At this point, information cannot 
propagate back to the Sun. Even if a field line ``closes'' beyond $12 R_\odot$, 
its apex inevitably will be convected outward by the solar wind to the outer 
domain boundary. Therefore, closed field is defined as those field lines with 
both footpoints at $1 R_\odot$ and an apex that does not reach $12 R_\odot$. 

As is evident in the top panels of Figure \ref{figure3}, the boundary before 
driving is mostly undisturbed and roughly
straight in both the north and south, except for the small 
ripples due to the solar wind and the non-steady 
background dynamics at the top of the helmet streamer. In the middle panels
of Figure \ref{figure3}, the northern boundary has been twisted by the
flow (see \S \ref{driver}) by $\pi/2$, whereas the southern boundary,
which was not driven, remains undisturbed from its initial state. 
The twisting caused by the photospheric motion is much 
larger than any of the ripples in the boundary caused by the non-steady 
background dynamics. By simply viewing Figure \ref{figure3}, we can conclude 
that the horizontal, open-closed boundary of the top panel has been rotated 
by approximately $\pi/2$ in the middle panel.  Moreover, the open-closed 
boundary appears to have undergone almost no relaxation during the 2.4 
\hrs \;of driving: it simply advects ideally with the photospheric flow. 
This is a key result with important implications for understanding the 
open-closed dynamics. The imposed photospheric flows produce a substantial 
deformation of the northern open-closed separatrix surface very early in 
the driving, and launch nonlinear Alfv\'en waves in both the open and closed 
fluxes. In principle, this could lead to the formation of current sheets and 
interchange reconnection along the separatrix or open-open reconnection once 
the wave reaches the HCS above the closed flux. However, we do not see any
evidence for such reconnection. The southern coronal-hole boundary appears
unchanged between the top and middle panels. 

It appears, therefore, that a significant relaxation requires the
buildup of a substantial deformation of the closed flux. This result
validates our arguments above that only long-time-scale photospheric
motions, such as supergranular flows, are important for driving the 
open-closed dynamics. For the parameters of our system, the typical
length of the last closed field line is $\sim 10^6$ km, whereas the
average Alfv\'en speed along this flux is $\sim 100$ \kms, so the 
Alfv\'en travel time is of order a few hours, longer than the 2.4
\hrs \;driving time. Consistent with this estimate is the result that,
in the simulation, electric currents due to the driving appear at the
southern footpoints of the rotated flux 3 \hrs \;after the start of the
driving. We expect, therefore, that the time scale for the decay of
the boundary deformation will be, at least, 3 \hrs. Any driving on
time scales much shorter than this, such as granules or the magnetic
carpet, will only add high-frequency noise to the coronal-hole boundary.

Figures \ref{figure3} and \ref{figure4} show this slow decay in detail.
Because we do not allow slipping on the photospheric surface after the
driving has ended, any change in the coronal-hole boundary is due to 1) 
opening of closed field, 2) closing of open field, or 3) interchange
reconnection between the two. Opening of magnetic field lines would 
register as black changing to white in Figures \ref{figure3} and 
\ref{figure4}, and corresponds to a closed field line rising up and 
moving past $R = 12 R_\odot$. Closing of magnetic field lines would 
register as white changing to black, and corresponds to two open 
field lines reconnecting to form a closed field line and a u-loop 
disconnected field line, which would move with the solar wind and 
leave through the outer boundary at 30 $R_\odot$. Finally, interchange 
reconnection would register as either a black-to-white or white-to-black 
switch, and occurs when one open and one closed field line reconnect, 
most likely in the vicinity of the HCS high in the corona, and 
switch foot points. Such an interaction results in the same amount of open 
and closed flux, but changes the connectivity of the system, allowing 
material that was trapped on the closed field line to move outwards 
into the solar wind. Note that the connectivity shown in all of these 
plots is instantaneous: any opening of a closed field line shown in 
the north would also appear immediately as opening in the south.

The bottom panel of Figure \ref{figure3} shows the coronal-hole
boundaries at $t = 11.6$ \hrs. By this time, the closed field has 
had sufficient time to 
deform substantially in response to the applied photospheric stress. 
The open-closed boundary in the north shows clear signs of activity, 
with an intrusion of open field cutting into the closed field 
near the location of the center of the flow. Sharp structure has also 
appeared on the boundary near the edge of the displaced region around 
$8^\circ$ longitude. The southern boundary, in contrast, remains smooth 
with no discernible changes.

The top panels of Figure \ref{figure4} show the system at $t = 21.0$ \hrs. 
At this time, the Alfv\'en wave on the open field lines has left the system 
through the outer boundary at $30 R_\odot$. The coronal-hole boundary has 
smoothed out somewhat and is beginning to show clear counter-rotation back 
towards its initial state. The bottom panels of Figure \ref{figure4} show 
time $t = 30.3$ \hrs, where the southern boundary remains largely unchanged. 
There is a small distortion near $-5^\circ$ longitude due to the 
non-steady background dynamics occurring at the top of the 
helmet streamer. The coronal-hole boundary in the north, on the other hand, 
has continued to evolve much more dramatically, counter-rotating 
back towards its initial, smooth configuration. 

The question now becomes: how is the previously displaced open magnetic 
field becoming closed again, and the displaced closed field becoming open? 
Any true opening or closing of the magnetic field would produce a signature 
in the south, yet the southern boundary remains largely unaltered. We must 
conclude, therefore, that interchange reconnection is responsible for the 
change in the northern boundary. Because this boundary is 
merely the 2D imprint of a 3D surface that extends well up into the corona 
and inner heliosphere, the interchange reconnection could be occurring 
anywhere within, or adjacent to, that 3D surface.
%, \textcolor{red}{although from only a 2D projection
%of a 3D boundary, we cannot yet say at exactly what height this 
%interchange reconnection has taken place}.

Figure \ref{figure5} shows the amount of interchange reconnection
experienced by the field lines traced in Figures \ref{figure3} and
\ref{figure4} in the northern (red) and southern (blue) hemispheres,
along with the amount of flux closing down (yellow) and opening up 
(green).  Of course, the amounts of opening and closing are identical 
in the two hemispheres, but the interchange can be very different. 
Figure \ref{figure5} begins at the end of the driving phase; consequently, 
any change in connectivity must be due to opening, closing, or reconnection. 
There is a continual, small, but measurable amount of closing throughout 
the simulation. This is due to an overall continual closing experienced 
by the system as it relaxes gradually from its initial PFSS state toward 
a steady configuration supported by the solar wind. In addition, a 
quasi-steady amount of interchange reconnection takes place in both 
hemispheres. This is due to the persistent restructuring of the HCS 
due to the non-steady background dynamics. Over and above 
these two processes, a transient enhancement of interchange reconnection occurs 
principally in the northern hemisphere, where the rate reaches a maximum 
about 3 \hrs \;after the end of the driving phase.  This is consistent with 
our argument above that the relaxation is driven by the deformation of 
the closed flux and, hence, reflects its intrinsic time scale. The decay 
time for this interchange is also of order a few hours. There appears to 
be a rise in interchange reconnection in the south, as well, but this is 
far less clear than in the north. It may well be that the deformation of 
the closed flux drives interchange at foot points that were not driven by 
the flow.

Our result that interchange reconnection is the dominant relaxation
process clarifies why a substantial deformation of the closed flux
must build up in order for the relaxation to occur. Interchange
reconnection cannot occur if the top of the helmet streamer maintains
its classic 2D geometry, with a simple Y-type null and a current sheet
only above the closed flux. This region must become strongly 3D, with
current sheets forming between open and closed flux \citep{wang00}. 
Since any stress on the open flux will simply propagate 
away, it must be the deformation of the closed flux that leads to the
current sheets and the ensuing interchange reconnection.  A somewhat
surprising result is the clear lack of significant field-line opening
during the relaxation. There is some weak opening before and during the 
driving, but this is negligible.  A seemingly obvious evolution for 
relaxing the field would be to open all the stressed closed flux and 
then simply close down all the flux that is not open in the original, 
pre-driven configuration.  This would return the system back to its 
minimum-energy state. In spite of its effectiveness, however, we see 
no evidence for such evolution. A possible explanation is that closed 
field lines do expand outward and attempt to open, but then encounter 
open flux and interchange reconnect before reaching $12 R_\odot$. In 
any case, our simulation clearly shows that a localized deformation 
of the open-closed boundary relaxes almost exclusively via interchange 
reconnection.

To better understand how interchange is able to produce a global
relaxation of the boundary back toward its original shape, we track 
where the interchange is occurring along the boundary. Figure \ref{figure6} 
shows the instantaneous change in connectivity on the left, and the 
accumulated change in connectivity on the right at the final time, 
$t = 30.3$ \hrs. (The full movie is available online.) 
%\textcolor{red}{We again are looking at merely the 2D imprint on the surface of the
%3D boundary, and so this plot can tell us nothing about where the dynamics take place radially.} 
Here again, we are viewing merely the 2D imprint of an 
extended 3D surface, and the governing interchange dynamics may be occurring 
anywhere along or very near that surface. The plot on the left showing the 
instantaneous change is a binary plot, where field lines that changed 
either from closed to open or from open to closed between the current 
and previous times are represented by red circles. The right panel 
shows how many times each field line changed connectivity throughout 
the simulation, where only field lines that have interchanged twice 
or more are shown. This means that any change in connectivity due to 
closing or opening in the overall PFSS relaxation would not be shown.

The movie online shows that, just as in the final frame shown here,
there are changes in connectivity all along the coronal-hole boundary
from $-20^\circ$ to $+20^\circ$ longitude, even though only the region 
within $\pm 7^\circ$ was displaced. From the right panel in Figure
\ref{figure6}, we can see that many field lines on the boundary interchange
more than 5 times, while some field lines in the region near the center of
the photospheric rotation interchange over 50 times. This result indicates 
that interchange reconnection must be the natural response to footpoint 
stressing, and that it must be common in the solar corona and wind.

We find that, through this interchange process occurring everywhere 
longitudinally along the boundary, locations far from 
the driven region in longitude are permanently changed. 
Figures \ref{figure7} and \ref{figure8} show the same closed field lines 
as displayed previously, except that here the field lines are colored by 
the locations of their southern foot points. Field lines with foot points 
between $-20^\circ$ and $-10^\circ$ longitude in the south are colored 
navy blue, those between $-10^\circ$ and $0^\circ$ are teal, those 
between $0^\circ$ and $+10^\circ$ are red, and those between $+10^\circ$ 
and $+20^\circ$ are yellow. In the initial field, these also correspond 
to the location of the northern foot points, since the field was potential 
with an axisymmetric mapping from south to north. Figures \ref{figure7} 
and \ref{figure8} allow us to observe clearly the change in the global 
mapping introduced by both the driver and the interchange reconnection. 
The top, middle, and bottom panels of Figure \ref{figure7} display the 
boundary at $t = -0.03$ \hrs \;before the driving, $t = 2.4$ \hrs \;immediately 
after the driving, and $t = 7.0$ \hrs, respectively.  Figure \ref{figure8} 
shows the late-time maps at $t = 16.3$ \hrs \;and $t = 30.3$ \hrs.

Our driver displaces field lines between $\pm 7^\circ$ longitude (compare 
the top and middle panels of Figure \ref{figure7}). However, we see in the 
other three panels of Figures \ref{figure7} and \ref{figure8} that field 
outside of this region is affected. In the bottom panel of Figure \ref{figure7}, 
the boundaries at $\pm 10^\circ$ have already begun to change, and still larger 
changes are evident in Figure \ref{figure8}. Most importantly, we see in the 
bottom panel of \ref{figure8} that the closed, teal-shaded flux that was the 
most displaced between $-5^\circ$ and $0^\circ$ has migrated down and pushed 
out the closed-field boundary between $-13^\circ$ and $-7^\circ$. We also see 
that yellow flux from beyond $+10^\circ$ has been displaced into the initially 
red region, even though this flux was not driven.

It is evident by comparing the first panel of  Figure \ref{figure7} to the 
last one in Figure \ref{figure8} that the system cannot return back to its 
original state, even if allowed to relax indefinitely. Figure \ref{figure5} 
shows that the amount of reconnection occurring at the end of the simulation 
has leveled off. Consequently, we conclude that by $t = 30.3$ \hrs \;the system
has reached a new quasi-steady state. This state differs from the original 
primarily in the presence of twist deep within the closed-field region. Since 
this twist is large-scale and far from the open-closed boundary, it does not 
produce any current sheets and cannot relax via reconnection. Even if reconnection 
could easily occur in the closed region, it could not return the field to its 
original, unstressed state, due to helicity conservation \citep[e.g.][]{taylor74, 
taylor86, antiochos13}. Of course, on the Sun a coronal-hole boundary never 
relaxes to some unperturbed state, because it is being driven continuously 
by the random supergranular flows. The key conclusion from the results presented 
here is that the time scale for full relaxation via interchange reconnection 
is of order 10 \hrs \;or so, which is comparable to the driving time. Therefore, 
solar coronal-hole boundaries are never quasi-steady, but are strongly dynamic.  

%%%%%%%%%%%%%%%%%%%%%%%%%%%%%%%%%%%%%%%%%%%%%%%%%%%%%%%%%%%%%%%%%%%%%%%%%%%%
\subsection{$2\pi$ Displacement}
%%%%%%%%%%%%%%%%%%%%%%%%%%%%%%%%%%%%%%%%%%%%%%%%%%%%%%%%%%%%%%%%%%%%%%%%%%%%

In order to determine the coronal-hole boundary dynamics for a strong,
continuous driver, we performed and analyzed a simulation with a full 
$2\pi$ rotation, corresponding to the complete lifetime of a supergranule. 
We kept the maximum velocity the same as in the $\pi/2$ case and drove the 
system at this steady rate for a longer duration, in order to reach the 
specified displacement. Because the top of the helmet streamer becomes 
much more distorted in this case, the results presented below exhibit 
much more drastic dynamics than those above. On the Sun, convective cells appear and disappear randomly 
on the photosphere continuously. The results presented below, therefore, 
still represent a great simplification from the true complexity of solar 
coronal-hole boundary dynamics.

Figures \ref{figure9} and \ref{figure10} show the coronal-hole boundary 
throughout the rotation.  As in Figures \ref{figure3} and \ref{figure4}, 
open field is represented by white and closed field by black, with the 
northern hemisphere shown on the left and the southern on the right. 
The top panels of Figure \ref{figure9} show the boundaries before onset 
of driving at $t = -0.03$ \hrs, and the middle panels show the boundaries
immediately after the driving has ceased at $t = 6.0$ \hrs. The grid has
sufficiently resolved one complete rotation of the coronal-hole boundary. 
While the boundary is strongly twisted in the north, there have been no 
changes to the southern boundary even though there has been sufficient 
time for the stress to propagate to the southern foot points. Note that 
the full $2\pi$ rotation induces an extreme deformation of the boundary, 
yielding much more fine structure than the $\pi/2$ case above. This is 
the reason for first analyzing and gaining insight from the case with 
a small rotation.  Observe also that the structure of the boundary at 
$t = 6.0$ \hrs \;is not due solely to ideal convection. There must have 
been considerable reconnection to form the detailed small-scale structure 
visible at the end of the driving phase. This is to be expected, given 
our finding above that the time scale for relaxation is 3 \hrs \;or so.

In light of this fine-scale structuring, it may seem surprising that 
even with the extreme deformation of the northern boundary, there is 
no observable effect on the southern. One reason for this result is 
that, on the time scale of 6 \hrs, any change in the south could be 
due only to interchange or closing. As noted previously, we define a 
field line to be open if it intersects the $R = 12 R_{\odot}$ surface. 
Since it takes approximately 12 \hrs \;or so in our simulation for a 
disturbance, whether mass flow or Alfv\'en wave, to propagate out to 
this surface, no opening can occur until later. The 
closing of field lines could, in principle, occur on the time scale 
of 6 \hrs, but this is unlikely given that the addition of large 
magnetic stress into the corona should result in a net opening of 
flux rather than closing. Furthermore, it is unlikely that a large 
amount of interchange would occur in the south, because this would 
only deform the boundary away from its initial quasi-steady position.

The bottom panels of Figure \ref{figure9} show the boundary at $t = 15.9$ 
\hrs, while Figure \ref{figure10} shows the boundaries at $t = 30.9$ \hrs, 
$t = 50.9$ \hrs, and $t = 75.6$ \hrs. These frames were chosen to best 
illustrate the phases of evolution. At early times, comparing $t = 6.0$
\hrs \;and $t = 15.9$ \hrs, we observe rapid evolution in the regions that 
are displaced the most. The movie available online shows the development 
and decay of fine-scale structure at the northern boundary throughout 
this process, just as in the $\pi/2$ case. Notice, however, that the 
evolution sometimes demonstrates the development of very fine corridors 
of both open and closed flux, but these corridors eventually disappear. 
The southern boundary, in contrast, remains unchanged during this time, 
despite the extreme dynamics in the north. By $t = 30.9$ \hrs \;(top 
panels of Figure \ref{figure10}), the changes in the north have slowed 
as the finest-scale structure has been interchanged away. We continue 
to see slow changes in the northern hemisphere along the lines of the 
behavior of the $\pi/2$ case. However, now the southern boundary has 
begun to show changes. Note in particular the spike-like indentation 
of open flux into the southern closed-field region. This feature could 
be due to opening of flux that has reached the 12 $R_{\odot}$ surface, 
but we will show below that it is due to interchange. Additionally, we 
see a distinct corrugation of both the northern and southern boundaries. 
This is due to localized opening or closing of flux, which occurs on a 
smaller scale even without the driving, but here it is enhanced by the 
induced stresses. The corrugation disappears over the next 10 \hrs \;or 
so. By the end of the simulation, which is nearly three days after the 
driving has stopped, the northern coronal-hole boundary has nearly, but 
not quite, returned to its initial state. The only significant change 
in the south, meanwhile, is that the spike of open flux has broadened 
into a smooth indentation, indicating that there has been a localized 
opening due to the stress that now resides permanently in the closed-field 
region.

Figure \ref{figure11} shows the amount of interchange reconnection,
closing, and opening that these field lines experience after the end 
of the driving, just as in Figure \ref{figure5}. As before, the red 
and blue curves indicate interchange reconnection in the north and 
south, and the yellow and green curves show the amount of closing 
and opening, respectively. Before the onset of driving, the amounts 
of interchange reconnection in the north and south match those found
in the $\pi/2$ case. As the driving ceases, however, the interchange
in the north increases rapidly, reaching a maximum rate approximately 
3 \hrs \;after the end of the driving. This time scale is the same as
found in the $\pi/2$ case, but the amplitude is 6 times larger here
for the larger displacement. As before, for this case also there is
no noticeable increase in the interchange rate in the south, at least 
on the scale of the figure.

The interchange in the south subsequently does show a broad rise and 
attains a maximum, but only about 20 \hrs \;after the end of the driving. 
In fact, the interchange rate in the south surpasses that in the north 
during this period. Notice that after about 40 \hrs \;or so, the interchange 
in the north again rises and eventually surpasses that in the south. It 
appears that there is a global oscillation in the interaction between 
open and closed flux during this relaxation. The stress in the closed 
field propagates quickly, at the Alfv\'en speed, and for a force-free 
field should equilibrate along field lines. We expect, therefore, that 
after 10 \hrs \;or so, the closed flux should reach an equilibrium. At 
first, this equilibrium produces current sheets in the open flux only 
in the north, which has been deformed by the flow. The relaxation in 
the north by interchange is sufficiently fast, however, that eventually 
it throws the south out of balance. The south then begins to interchange 
rapidly as well, and surpasses that occurring in the north. This exchange 
occurs once more before the two rates decline to become almost identical 
by $t = 65$ \hrs. Although interesting, this oscillation is not relevant 
to the Sun, where both the northern and southern boundaries would be 
driven simultaneously and continuously. It does, however, 
emphasize the 3D nature of the dynamics. It also hints that the location 
experiencing interchange reconnection is more likely to be near the top 
of the helmet streamer than close to the solar surface, as this location 
migrates over time from the northern to the southern hemisphere and back 
again. We discuss the radial location of the interchange reconnection in 
more detail in \S \ref{discussionsection}.

Figure \ref{figure12} is the final frame of the available online movie, 
which shows the instantaneous and accumulated changes in connectivity 
due principally to interchange reconnection. The left side shows that, 
as before, interchange reconnection is occurring at all 
longitudes along the entire 
coronal-hole boundary. The right shows that hot spots of connectivity 
changes exist in regions with the most structure, and that individual 
field lines interchange over 50 times, as observed earlier. Note that 
even in the south, where there is no photospheric driving, some 
locations show numerous interchanges taking place.

The long-term effect of this large amount of interchange on its
surroundings is shown in Figures \ref{figure13} and \ref{figure14}, 
in the same style as Figures \ref{figure7} and \ref{figure8}. Closed 
field lines again are grouped into four regions, based on the locations 
of their southern footpoints.  The top panel of Figure \ref{figure13} 
shows the corona-hole boundary and closed-field region before driving 
at $t = -0.03$ \hrs, the middle immediately after driving at $t = 6.0$ 
\hrs, and the bottom at $t = 15.9$ \hrs. Figure \ref{figure14} shows 
late-time snapshots at $t = 30.9$ \hrs \;and $t = 75.6$ \hrs. (The 
full movie is available online.)

The interchange reconnection works quickly at the beginning. At $t = 
15.9$ \hrs \;(bottom panel of Figure \ref{figure13}) field lines in the 
teal and red regions that were displaced can already be seen cascading 
along the boundary toward $-10^\circ$ longitude. In the top panel of 
Figure \ref{figure14}, the interchange continues to push the boundary 
north around $-10^\circ$ longitude. Also, a teal-shaded finger extends
all of the way to $-15^\circ$ longitude, $6^\circ$ beyond the initial
flow region, while field lines from the yellow-shaded region beyond 
$+10^\circ$ longitude have interchanged their way along a large portion 
of the boundary, despite not having been driven. Notice also that there 
is clear evidence for reconnection within the closed-field region. A 
long yellow-shaded filament, which at one time extends far into the red 
region, retracts considerably back toward its original location. The 
bottom panel of Figure \ref{figure14} at time $t = 75.6$ \hrs, nearly 
three days after the end of the driving, shows that even though the 
boundary may appear smooth and close to its original configuration, 
the closed-field region remains quite far from its initial state. 
Of course, any such structure would not survive long on the Sun, 
because subsequent photospheric rotations would completely destroy it.

%%%%%%%%%%%%%%%%%%%%%%%%%%%%%%%%%%%%%%%%%%%%%%%%%%%%%%%%%%%%%%%%%%%%%%%%%%%%
\section{Discussion}\label{discussionsection}
%%%%%%%%%%%%%%%%%%%%%%%%%%%%%%%%%%%%%%%%%%%%%%%%%%%%%%%%%%%%%%%%%%%%%%%%%%%%

The key result of our simulations is that interchange reconnection
dominates the evolution of the open-closed boundary when driven by a
photospheric rotation. Although some flux opening and closing does
occur, it is far outweighed by interchange. As discussed above, this
is somewhat unexpected, because flux opening is generally assumed to
be the mechanism by which the corona sheds magnetic stress, certainly
the large-scale magnetic shear evident in CMEs \citep[e.g.][]{lynch16}.
Interchange reconnection is typically not expected to produce marked
changes in coronal-hole boundaries, since interchange at any arbitrary 
location along a perturbed boundary is not guaranteed to help the 
boundary evolve back toward its lowest-energy state. In Figures 
\ref{figure3} and \ref{figure4}, interchange would show up as a 
switching of a black and white point, so if this occurs across 
the boundary at the point of maximum displacement, the interchange 
actually could move the system farther from its preferred state. 
Instead, the tip of the displaced closed-field region interchanges 
with field lines along the boundary in a manner that broadens the 
closed-field region but smoothes the boundary. The result is displaced 
field lines pushing along the coronal-hole boundary in a manner that 
makes the zone of maximum displacement appear to be diffusing back 
into the closed-field region. This effect can be seen in Figure \ref{figure8}, 
where the teal-shaded region of field lines is pushing into the navy 
and red regions.

The key to understanding this evolution is that interchange reconnection 
usually acts to smooth out any sharp structure that forms along the boundary. 
We emphasize, however, that interchange does not always smooth out structure.  
The movies of the $2\pi$ case clearly show the transient formation of long, 
thin filaments of both closed and open field during the relaxation. Furthermore,
the southern boundary often builds up fine-scale structure, even though this 
boundary was not driven. The structures due to interchange are fairly short-lived, 
but they do show that interchange cannot be thought of as a purely diffusive 
process.  For the most part, however, interchange reconnection evolves the 
entire boundary, not just where the displacement occurs, so as to smooth 
out any sharp features. Our simulations show that localized photospheric 
driving produces a permanent change in the open-closed boundary, even far 
from the driven region.

A heuristic picture of the effects of interchange reconnection is shown in 
Figure \ref{figure15}. Panels a, b, and c illustrate how 
the coronal-hole boundary 
(solid black line) changes with time. The closed-field region is represented 
by the entire solid-shaded gray, red, blue, orange, and green regions beneath 
the solid black line. The striped regions above the solid black line are 
open-field regions. Between panels a and b, interchange reconnection occurs 
between the closed-field region 1 and open-field region 2, and between the
closed-field region 3 and open-field region 4. The pattern repeats on the 
left side, with like-colored regions interchanging. The new boundary that 
results from this interchange is shown as the yellow line in panel a and 
becomes the new coronal-hole boundary shown in black in panel b. The 
process then continues, resulting in the black coronal-hole boundary 
shown in panel c. The original boundary is shown in gray, over the top 
of the new one in black, in panel c. The overall effect is to shrink the
height of the region while broadening it and preserving its area. A large 
fraction (though not all) of the evolution seen in our simulations is 
simply the cumulative effect of this process.

Two important aspects of the evolution of our system are implied by Figure 
\ref{figure15}, but may not be immediately apparent. First, if we were to 
repeat the process shown in panels a and b on the black triangle of panel 
c, then a section of the closed-field region lower down along each side, 
which previously was open, would re-open. Some of this section could close 
again in a subsequent set of interchanges. This explains the result of 
Figure \ref{figure9}, which shows that some locations undergo numerous 
interchange reconnections. Second, note that the center of photospheric 
rotation is likely to be a location of many interchanges. The interchange
process illustrated in Figure \ref{figure15}, occurring about the center 
of the boundary deformation for the $\pi/2$ case of Figure \ref{figure3}, 
would act very much like a counter-rotation of the boundary. It is especially 
advantageous for interchange to occur here and, thereby, to bring the system 
back to its original state. Figure \ref{figure6} demonstrates that, in fact, 
interchange occurs around this location over 50 times in 30 \hrs.

Figure \ref{figure16} presents an example of interchange reconnection that 
occurs within our system at approximately $t = 13$ \hrs. It shows a set of 
field lines drawn from fixed locations in the northern hemisphere at two 
different times, five minutes apart, during the $2\pi$ simulation. The 
field line in red is traced from exactly the same location in both images. 
In the top panel, the red field line is closed, and in the bottom, it is 
open. The interchange-reconnection point is at the sharp kink shown in 
the bottom panel, separating the old part of the field line, which obviously 
has not changed as it is traced back towards the Sun, from the new part of
the field line with its open end. This kink verifies that the interchange
reconnection is occurring principally at the top of the helmet streamer. 
The fact that the interchange takes place near the streamer 
top is not surprising because, 
for a force-free field, any magnetic twist or stress is expected to 
propagate to the region of weakest field \citep{parker79}. 
In our case, this corresponds to the top of the helmet streamer. It 
also happens to be the location of a pre-existing current sheet, the 
HCS, enabling the formation and enhancement of current sheets. Plasma 
distributed along the field line from the point of the kink down to 
the surface is now free to flow outwards into the solar wind, after 
residing in the hot corona on a closed field line.

Our result that a single field line can interchange over 50 times
implies that a spacecraft sampling along a single field line could, 
in fact, sample plasma from regions that are remote from each other 
on the Sun. This would have the effect of spreading plasma from closed 
field lines out into the heliosphere, with the interchange reconnection 
acting as a kind of diffusion. The S-Web model \citep{antiochos11} 
predicts that photospheric motions broaden the regions where reconnection
is likely to occur and increase the width of the source regions of slow 
solar wind in the heliosphere. Figure \ref{figure17} shows the width
of the dynamic slow-wind region in our simulation along the HCS. On
the left, at time $t = -0.03$ \hrs, line-tied field lines exhibit the 
dipolar shape of the magnetic structure before onset of driving, while 
field lines plotted within the HCS show the tangled magnetic fields 
that are released by the pinching off of the helmet 
streamer and convected out with the solar wind. In our case, the dynamic 
region near the HCS results from the non-steady background 
fluctuations due to numerical reconnection, so that 
the width of this dynamic region is determined by the grid. On the right, 
at time $t = 29.9$ \hrs, the same blue line-tied field lines are plotted 
along with open field lines traced from within the flow region on the 
surface. Whereas on the left the dynamic region consists only of a narrow, 
5$^\circ$-wide region due to the non-steady background dynamics, 
on the right the width of the dynamic region is much broader,
about $15^\circ$. Open field lines within this region trace outwards as 
far as $15^\circ$, but they also wrap into the HCS, mixing plasma from 
these two very different sources.

%%%%%%%%%%%%%%%%%%%%%%%%%%%%%%%%%%%%%%%%%%%%%%%%%%%%%%%%%%%%%%%%%%%%%%%%%%%%
\section{Conclusions}\label{conclusions}
%%%%%%%%%%%%%%%%%%%%%%%%%%%%%%%%%%%%%%%%%%%%%%%%%%%%%%%%%%%%%%%%%%%%%%%%%%%%

The simulations presented in this paper reveal several important new 
properties of the open-closed field boundary in the solar corona and 
heliosphere. The first is that, on the Sun, the magnetic field near 
this boundary responds to photospheric driving primarily via interchange 
reconnection near the top of the helmet streamer. This 
result holds for the flows that are likely to be the 
most important drivers of the boundary: the quasi-random twisting and 
shearing, on time scales of a day or so, induced by the ever-present 
supergranular convection. Although we certainly see closed flux opening 
and open flux closing in our simulations, these processes are minor 
compared to the interchange reconnection. In fact, since interchange 
reconnection is so efficient, we expect that it also dominates the
coronal-hole boundary evolution due to other driving motions, such 
as differential rotation or meridional flow, which have much larger 
spatial and temporal scales. This result holds not 
only for more complex driving motions, but also for a more complex 
coronal-hole boundary. All that is required is for the coronal hole 
to be significantly larger than the scale of the driver and for it 
to be adjacent to a helmet streamer, thus connecting it directly to 
the heliospheric current sheet.

We conjecture that interchange will always dominate if the photospheric 
driving is fully 3D and if it involves mainly the displacement of existing 
photospheric flux, rather than a change in the amount of flux. If a large 
amount of new flux emerges so that the total unsigned flux at the photosphere 
increases significantly, then we expect that the amount of open flux must 
increase as well. Flux emergence, therefore, will result in closed flux 
opening, although there may also be some interchange. Conversely, photospheric 
flux disappearance will result in open flux closing. Large changes in the amount 
of photospheric flux, however, occur on the time scale of the solar cycle, of 
order years, whereas the time scale for flux evolution due to supergranular 
motions is only of order a day. We conclude, therefore, that interchange 
reconnection almost always dominates the dynamics at coronal-hole boundaries.

Another important finding from our simulations is that the time scale
for boundary relaxation by interchange reconnection is commensurate
with supergranular driving. It is evident from our $2\pi$ case that
at an average speed of $\sim 9$ \kms, the photospheric motion induces
a strong, nearly ideal, deformation of the coronal-hole boundary. Some
relaxation occurred over the 6 \hrs \;duration of our driver, but the
interchange dynamics peaked several hours after the end of the driving 
and were significant for at least 30 \hrs.  If we were to reduce the 
driver speed by a factor of 5, so as to more closely match the solar 
photosphere, we expect that the boundary deformation would be smaller, 
but there would still be a pronounced deformation of the boundary and 
copious interchange reconnection. A possible caveat to this conclusion 
is that the effective Lundquist number for our simulation, which we 
estimate to be $\sim 10^4$, is orders of magnitude smaller than the 
actual solar value, $\sim 10^{12}$.  In principle, this could result 
in much slower reconnection than is produced by our simulation and, 
thereby, alter the balance between interchange versus opening and 
closing.  However, there are compelling arguments against this 
possibility. A key point is that our system has a large-scale 
separatrix surface, the open-closed boundary, with a null line and 
an initial current sheet, the HCS. A large number of simulations, 
by ourselves and many others, of systems with such pre-existing 
separatrices and null points show that stressing such topologies 
leads to the formation of current sheets at the separatrices on 
ideal-MHD time scales \citep{priest00}. Furthermore, the 
rate of interchange reconnection need not be fast, but must 
merely keep pace with the slow photospheric driving. In our 
simulations, the reconnection rate is of order a few percent 
of the Alfv\'en speed, and there is no evidence of explosive 
reconnection. We expect, therefore, that on the Sun, where the 
driving is continuous and slow, the interchange reconnection 
would achieve a quasi-steady balance with the photospheric 
motions \citep[e.g.][]{edmondson2010a}.

Our results also have important implications for understanding the
topology of the open and closed flux in the corona. Careful examination 
of the movie for the $ 2\pi$ case reveals that, in spite of the extremely 
fine structure that develops during the evolution, both the closed- 
and open-flux regions remain simply connected. There is no evidence 
for disconnected open-flux patches within the closed-field region, 
even though the system is fully dynamic and includes numerous current 
sheets. This agrees with other simulations of coronal-hole boundary 
evolution \citep{edmondson10b, linker11} and supports the coronal-hole 
uniqueness hypothesis of \citet{antiochos07}. This result also provides 
indirect support for the S-Web model for slow-wind origin \citep{antiochos11}, 
in contrast to the interchange model of Fisk and co-workers, which postulates 
that open flux can diffuse throughout the closed-field region \citep{fisk98, fisk03}.

For understanding in-situ measurements of the slow wind, the most
important conclusion from our simulations is that magnetic flux near
the open-closed boundary is constantly undergoing cycles of opening
and closing via interchange near the inner edge of the 
heliospheric current sheet, on time scales of approximately 30 \hrs. 
This implies that closed-loop plasma that has been in the corona for 
a day or more is continuously being released into the solar wind. Such
plasma could well explain the observed characteristics of the slow
wind, a charge-state abundance indicative of the closed corona, and
significant FIP enhancements \citep{zurbuchen07}. We emphasize 
that the time scale of 30 \hrs \;is critical, because it takes roughly 
a day for coronal loops to build up an elemental abundance that differs 
significantly from the photosphere \citep{feldman02}. Our results also 
can explain the observation that most of the slow wind does not exhibit 
bi-directional heat fluxes, which are evidence of flux opening, or 
heat-flux dropouts, which are evidence of flux closing \citep{gosling90,lin92}. 
For remote-sensing observations of the corona, our calculations predict
that the boundaries of coronal holes should have highly irregular
structure, on scales considerably smaller than supergranules, and
should exhibit a slow, quasi-cyclic evolution. It may be possible 
to identify such an evolution from high-resolution images of 
coronal holes.

It must be noted, however, that the Sun exhibits several features that 
are not included in our simulations and are likely to have strong effects 
on any observations of the corona and wind. In particular, the photosphere 
displays persistent emergence and cancellation of magnetic flux at both 
small and large scales. We expect this to drive systematic evolution of 
coronal-hole boundaries, including large-scale opening and closing, while 
small-scale emergence/cancellation will produce a constant background of 
brightening and dimming in coronal images. Even more important, the flux 
distribution at the photosphere is never that of a simple dipole. As a 
result, the distribution of coronal holes is almost never that of two 
simple polar holes, as in our simulation. The coronal-hole structure 
as inferred by either source-surface or MHD models \citep{titov11} 
generally is intricately organized, with multiple coronal-hole extensions 
reaching low solar latitudes. According to the S-web model, such a complex 
coronal-hole topology is essential for understanding the observations that 
slow wind can be found far from the HCS \citep{antiochos07, antiochos11}. 
Future simulations with much higher numerical resolution will be required 
to understand the dynamical response of such a complex open-closed boundary 
to photospheric driving and to determine whether the ensuing dynamics can 
explain the slow solar wind.

%%%%%%%%%%%%%%%%%%%%%%%%%%%%%%%%%%%%%%%%%%%%%%%%%%%%%%%%%%%%%%%%%%%%%%%%%%%%
%%END TEXT
%%%%%%%%%%%%%%%%%%%%%%%%%%%%%%%%%%%%%%%%%%%%%%%%%%%%%%%%%%%%%%%%%%%%%%%%%%%%

A.K.H., S.K.A., and C.R.D.\ were supported by NASA Living With a Star and 
Heliophysics Supporting Research projects. P.F.W.\ acknowledges support 
from the NASA Postdoctoral Program at GSFC. The large, lengthy numerical 
simulations performed and analyzed in this paper were supported by grants 
of NASA High-End Computing resources to C.R.D.\ and were carried out on 
the discover cluster at NASA's Center for Climate Simulation.

%%%%%%%%%%%%%%%%%%%%%%%%%%%%%%%%%%%%%%%%%%%%%%%%%%%%%%%%%%%%%%%%%%%%%%%%%%%%%

%%%%%%%%%%%%%%%%%%%%%%%%%%%%%%%%%%%%%%%%%%%%%%%%%%%%%%%%%%%%%%%%%%%%%%%%%%%%
%%FIGURE1 - DOMAIN & GRID
%%%%%%%%%%%%%%%%%%%%%%%%%%%%%%%%%%%%%%%%%%%%%%%%%%%%%%%%%%%%%%%%%%%%%%%%%%%%

%%%%%%%%%%%%%%%%%%%%%%%%%%%%%%%%%%%%%%%%%%%%%%%%%%%%%%%%%%%%%%%%%%%%%%%%%%%%

%%%%%%%%%%%%%%%%%%%%%%%%%%%%%%%%%%%%%%%%%%%%%%%%%%%%%%%%%%%%%%%%%%%%%%%%%%%%
%%FIGURE 2 - ZOOM IN ON FLOW REGION W/IN FIGURE 1 
%%%%%%%%%%%%%%%%%%%%%%%%%%%%%%%%%%%%%%%%%%%%%%%%%%%%%%%%%%%%%%%%%%%%%%%%%%%%

%%%%%%%%%%%%%%%%%%%%%%%%%%%%%%%%%%%%%%%%%%%%%%%%%%%%%%%%%%%%%%%%%%%%%%%%%%%%

%%%%%%%%%%%%%%%%%%%%%%%%%%%%%%%%%%%%%%%%%%%%%%%%%%%%%%%%%%%%%%%%%%%%%%%%%%%%
%%FIGURE 3 - HALF PI - SNAPSHOTS OF BLACK AND WHITE BOUNDARY MOVIE (1/2)
%%%%%%%%%%%%%%%%%%%%%%%%%%%%%%%%%%%%%%%%%%%%%%%%%%%%%%%%%%%%%%%%%%%%%%%%%%%%

\begin{figure*}[!htbp]

\centering\includegraphics[width=\textwidth]{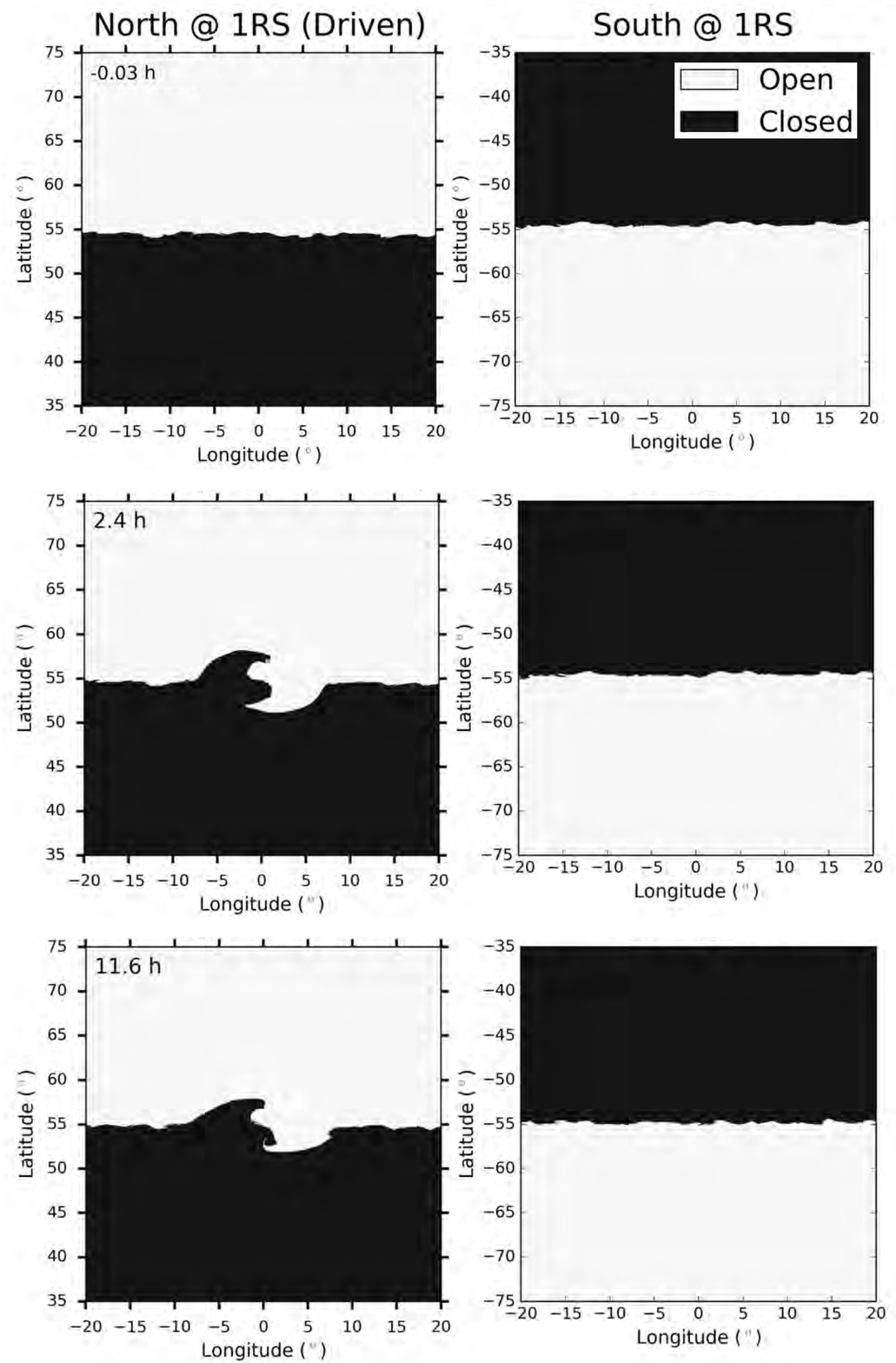}

\caption{Coronal-hole boundary maps for a peak displacement of $\pi/2$, in the north 
(left) and south (right). White is open magnetic field, and black is closed. Top: Before 
driving, at $t = -0.03$ \hrs. Middle: End of driving, at $t = 2.4$ \hrs. Bottom: At 
$t = 11.6$ \hrs. The full, 5-minute-cadence movie is available online.}

\label{figure3}
%%%%%%%%%%%%%%%%%%%%%%%%%%%%%%%%%%%%%%%%%%%%%%%%%%%%%%%%%%%%%%%%%%%%%%%%%%%%
\end{figure*}
%%%%%%%%%%%%%%%%%%%%%%%%%%%%%%%%%%%%%%%%%%%%%%%%%%%%%%%%%%%%%%%%%%%%%%%%%%%%

%%%%%%%%%%%%%%%%%%%%%%%%%%%%%%%%%%%%%%%%%%%%%%%%%%%%%%%%%%%%%%%%%%%%%%%%%%%%
%%FIGURE 4 - HALF PI - SNAPSHOTS OF BLACK AND WHITE BOUNDARY MOVIE (2/2)
%%%%%%%%%%%%%%%%%%%%%%%%%%%%%%%%%%%%%%%%%%%%%%%%%%%%%%%%%%%%%%%%%%%%%%%%%%%%

\begin{figure*}[!ht]

\centering\includegraphics[width=\textwidth]{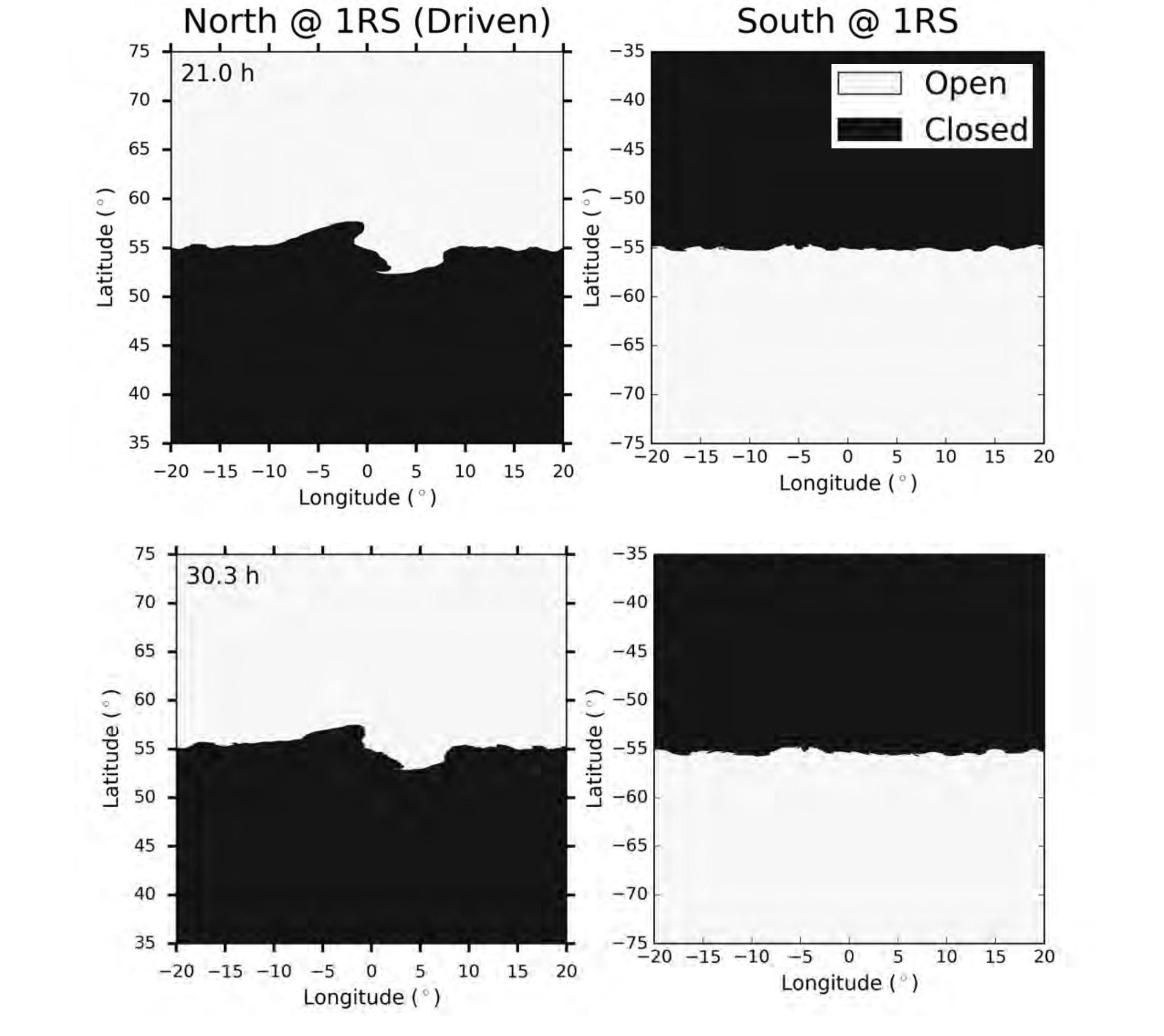}

\caption{Continuation of Figure \ref{figure3}. Top: At $t = 21.0$ \hrs. Bottom: At 
$t = 30.3$ \hrs. The full, 5-minute-cadence movie is available online.}

\label{figure4}
%%%%%%%%%%%%%%%%%%%%%%%%%%%%%%%%%%%%%%%%%%%%%%%%%%%%%%%%%%%%%%%%%%%%%%%%%%%%
\end{figure*}
%%%%%%%%%%%%%%%%%%%%%%%%%%%%%%%%%%%%%%%%%%%%%%%%%%%%%%%%%%%%%%%%%%%%%%%%%%%%

%%%%%%%%%%%%%%%%%%%%%%%%%%%%%%%%%%%%%%%%%%%%%%%%%%%%%%%%%%%%%%%%%%%%%%%%%%%%
%%FIGURE 5 - HALF PI - LINE PLOT OF INTERCHANGE VS CLOSING FLUX 
%%%%%%%%%%%%%%%%%%%%%%%%%%%%%%%%%%%%%%%%%%%%%%%%%%%%%%%%%%%%%%%%%%%%%%%%%%%%

\begin{figure*}[!htbp]

\centering\includegraphics[width=\textwidth]{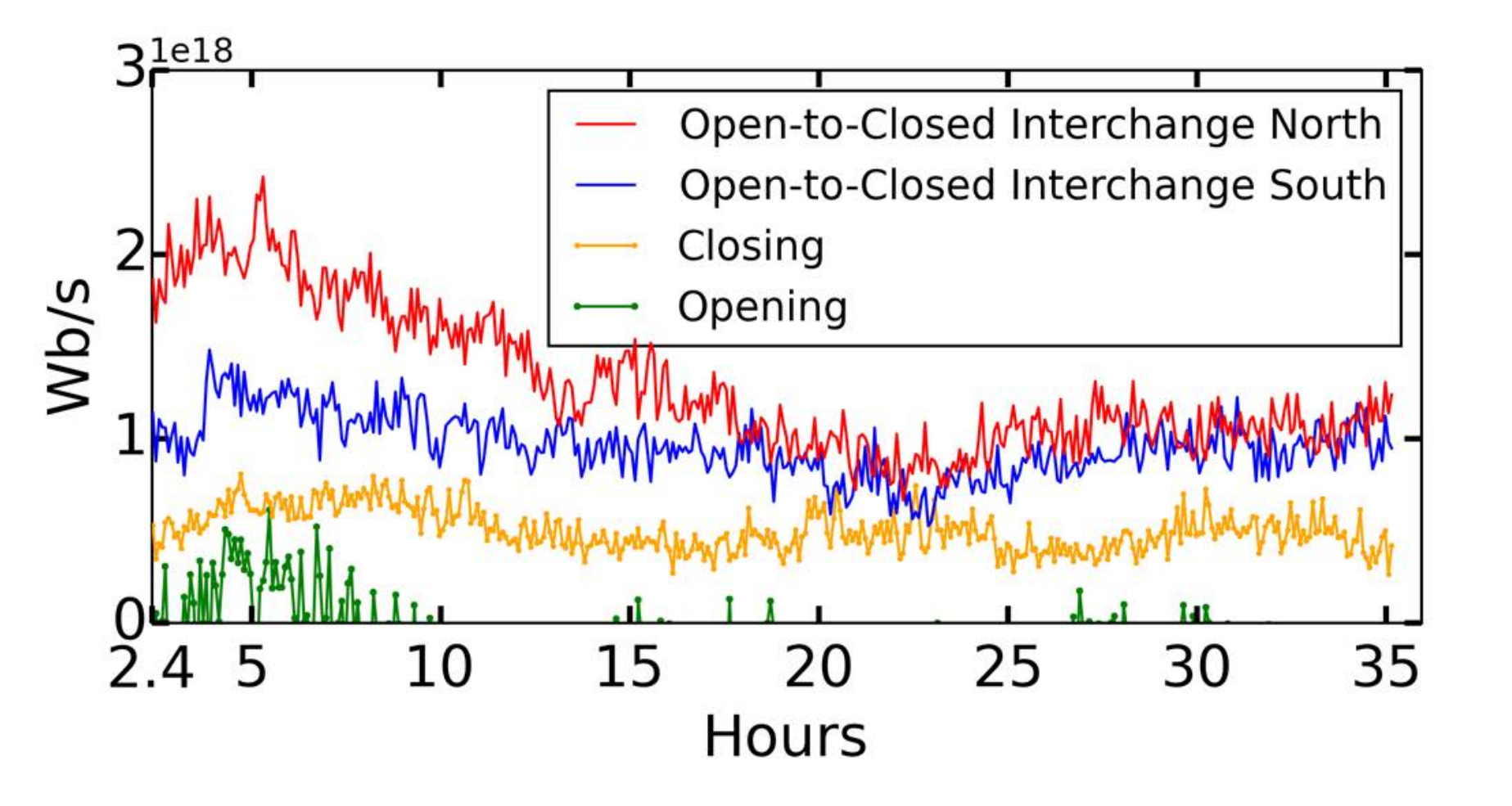}

\caption{Rate of flux change for a peak displacement of $\pi/2$, where time $t = 2.4$ \hrs \;
corresponds to the end of the driving. Red and blue curves display the rate of flux change 
due to interchange reconnection in the north and south, respectively; yellow and green curves 
display the rate of flux change due to closing and opening, respectively. The rates of opening 
and closing are identical in the north and south.}

\label{figure5}
%%%%%%%%%%%%%%%%%%%%%%%%%%%%%%%%%%%%%%%%%%%%%%%%%%%%%%%%%%%%%%%%%%%%%%%%%%%%
\end{figure*}
%%%%%%%%%%%%%%%%%%%%%%%%%%%%%%%%%%%%%%%%%%%%%%%%%%%%%%%%%%%%%%%%%%%%%%%%%%%%

%%%%%%%%%%%%%%%%%%%%%%%%%%%%%%%%%%%%%%%%%%%%%%%%%%%%%%%%%%%%%%%%%%%%%%%%%%%
%%FIGURE 6 - HALF PI - LAST FRAME OF ACCUMULATED INTERCHANGES PLOT 
%%%%%%%%%%%%%%%%%%%%%%%%%%%%%%%%%%%%%%%%%%%%%%%%%%%%%%%%%%%%%%%%%%%%%%%%%%%%

\begin{figure*}[!htbp]

\centering\includegraphics[width=\textwidth]{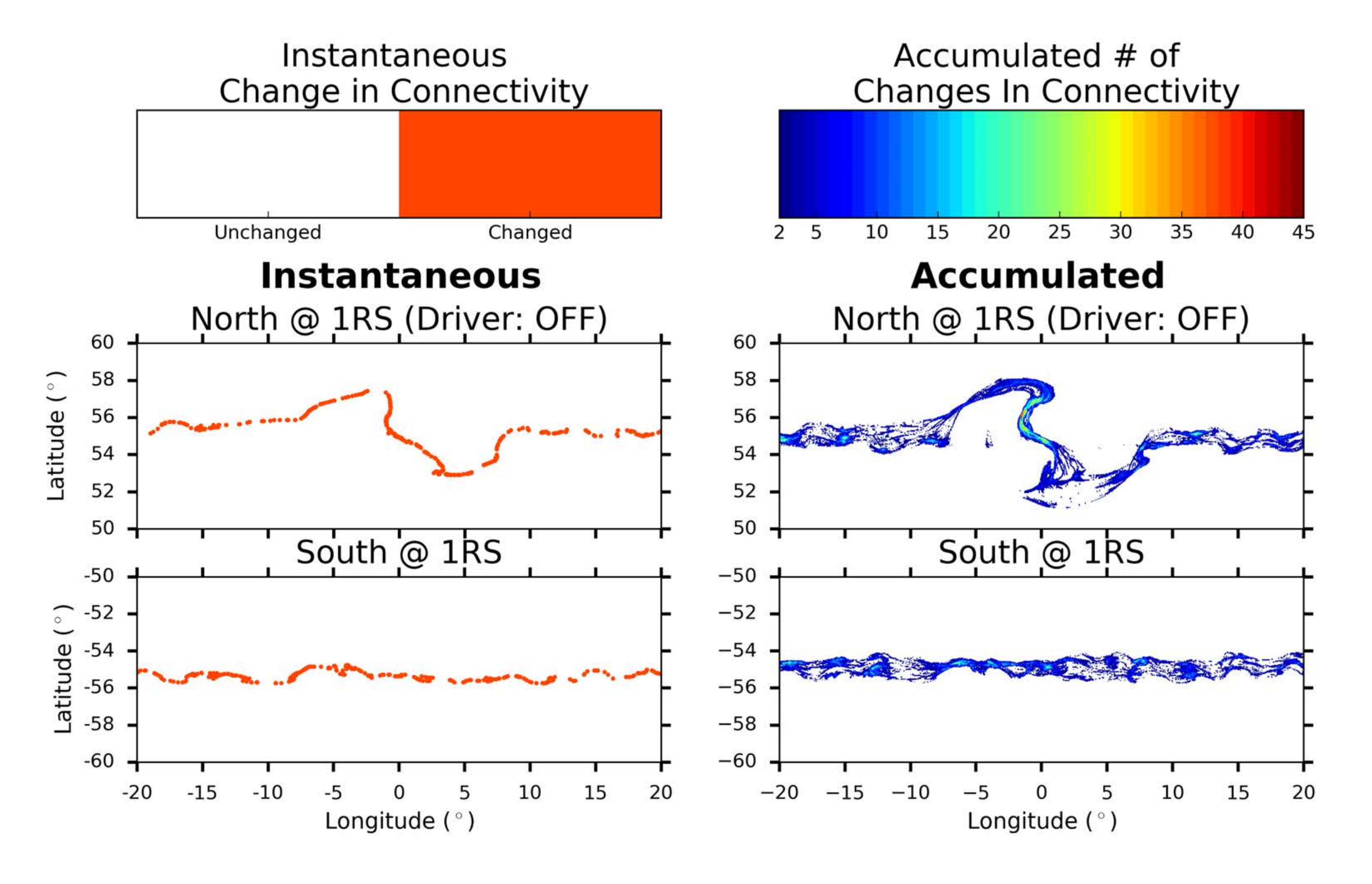}

\caption{Left: Instantaneous change in connectivity for a peak displacement of $\pi/2$, 
in the north (top) and south (bottom). Red circles indicate the locations of field lines 
that have changed their connectivity between the previous and current snapshot. Shown here 
are the final maps at $t = 30.3$ \hrs. Right: Contour plot of the number of times each field 
line has changed connectivity over the entire duration of the simulation, in the north (top) 
and south (bottom). Field lines that have changed connectivity only once are not shown. 
The full, 5-minute-cadence movie is available online.}

\label{figure6}
%%%%%%%%%%%%%%%%%%%%%%%%%%%%%%%%%%%%%%%%%%%%%%%%%%%%%%%%%%%%%%%%%%%%%%%%%%%%
\end{figure*}
%%%%%%%%%%%%%%%%%%%%%%%%%%%%%%%%%%%%%%%%%%%%%%%%%%%%%%%%%%%%%%%%%%%%%%%%%%%%

%%%%%%%%%%%%%%%%%%%%%%%%%%%%%%%%%%%%%%%%%%%%%%%%%%%%%%%%%%%%%%%%%%%%%%%%%%%%
%%FIGURE 7 - HALF PI - SNAPSHOTS OF FOUR COLOR DOT PLOT (1/2) 
%%%%%%%%%%%%%%%%%%%%%%%%%%%%%%%%%%%%%%%%%%%%%%%%%%%%%%%%%%%%%%%%%%%%%%%%%%%%

\begin{figure*}[!htbp]

\centering\includegraphics[width=\textwidth]{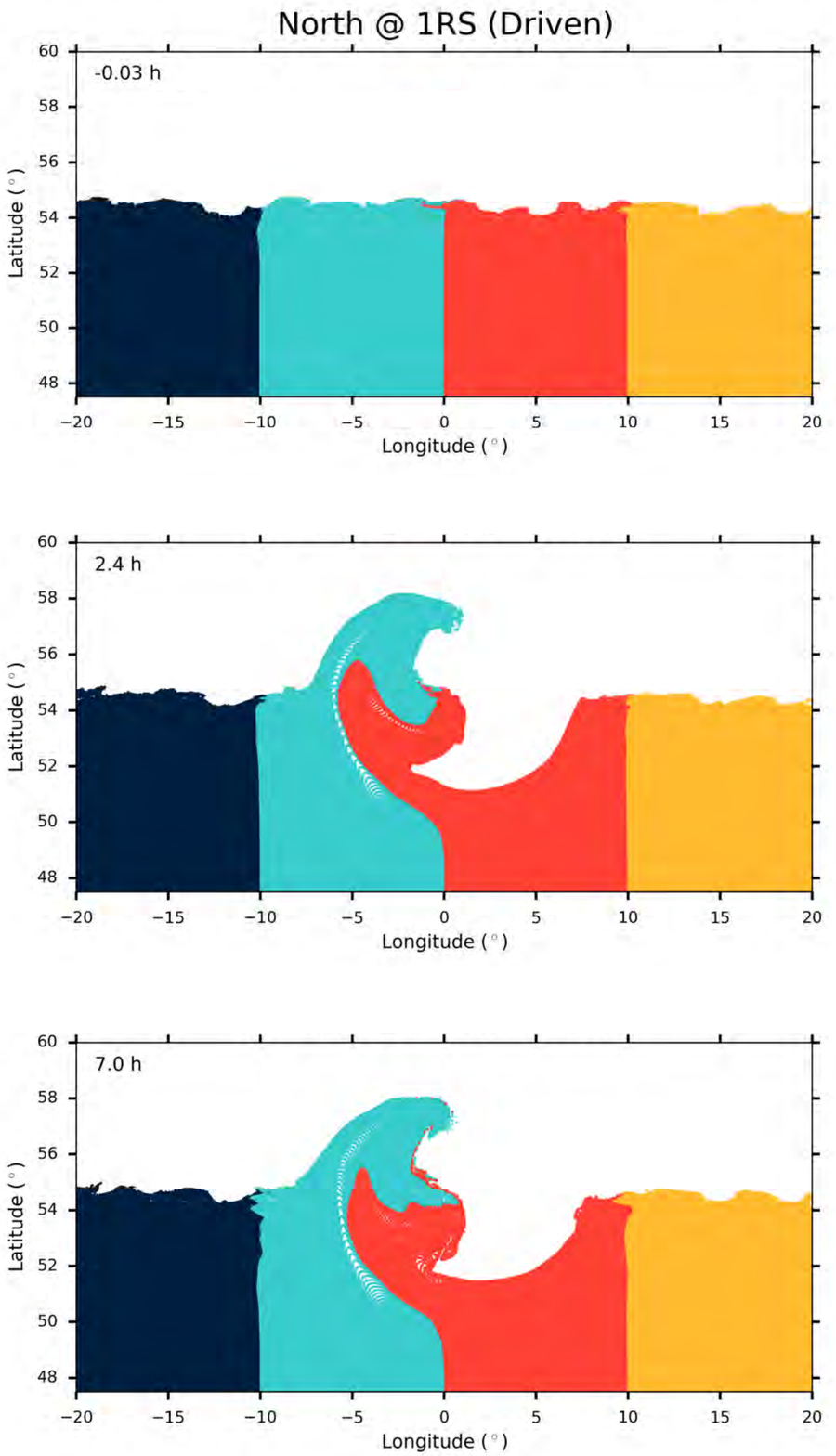}

\caption{Closed field lines in the north for a peak displacement of $\pi/2$. Field lines 
are color-shaded based on the longitudes of their conjugate footpoints in the southern 
hemisphere: navy blue between $-20^\circ$ and $-10^\circ$; teal between $-10^\circ$ and 
$0^\circ$; red between $0^\circ$ and $+10^\circ$; and yellow between $+10^\circ$ and 
$+20^\circ$. Top: Before driving at $t = -0.03$ \hrs. Middle: End of driving at $t = 
2.4$ \hrs. Bottom: At $t = 7.0$ \hrs. The full, 5-minute-cadence movie is available 
online.}

\label{figure7}
%%%%%%%%%%%%%%%%%%%%%%%%%%%%%%%%%%%%%%%%%%%%%%%%%%%%%%%%%%%%%%%%%%%%%%%%%%%%
\end{figure*}
%%%%%%%%%%%%%%%%%%%%%%%%%%%%%%%%%%%%%%%%%%%%%%%%%%%%%%%%%%%%%%%%%%%%%%%%%%%%

%%%%%%%%%%%%%%%%%%%%%%%%%%%%%%%%%%%%%%%%%%%%%%%%%%%%%%%%%%%%%%%%%%%%%%%%%%%%
%%FIGURE 8 - HALF PI - SNAPSHOTS OF FOUR COLOR DOT PLOT (2/2) 
%%%%%%%%%%%%%%%%%%%%%%%%%%%%%%%%%%%%%%%%%%%%%%%%%%%%%%%%%%%%%%%%%%%%%%%%%%%%

\begin{figure*}[!htbp]

\includegraphics[width=\textwidth]{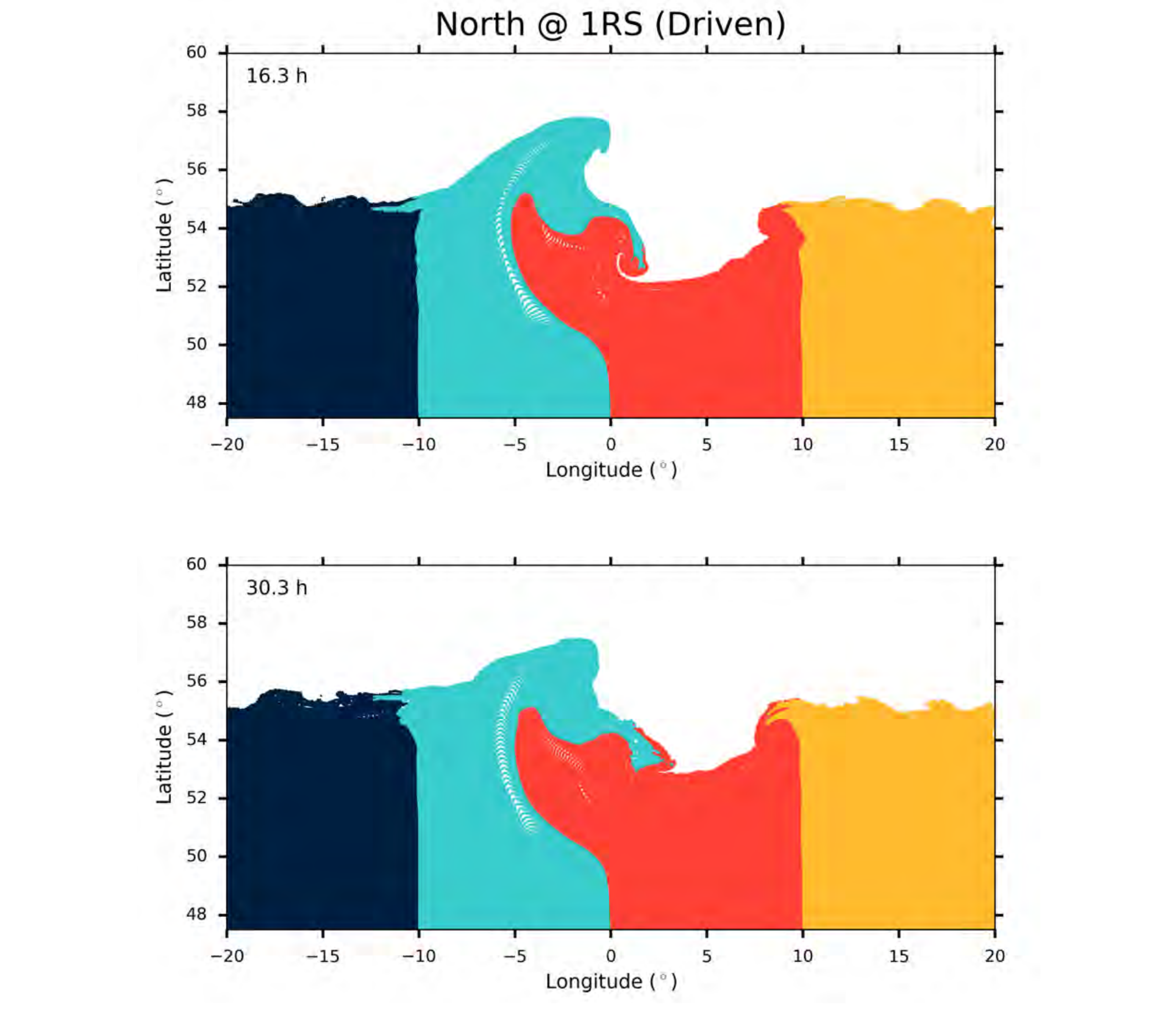}

\caption{Continuation of Figure \ref{figure7}, using the same color scheme. Top: At 
$t = 16.3$ \hrs. Bottom: At $t = 30.3$ \hrs. The full, 5-minute-cadence movie is 
available online.}

\label{figure8}
%%%%%%%%%%%%%%%%%%%%%%%%%%%%%%%%%%%%%%%%%%%%%%%%%%%%%%%%%%%%%%%%%%%%%%%%%%%%
\end{figure*}
%%%%%%%%%%%%%%%%%%%%%%%%%%%%%%%%%%%%%%%%%%%%%%%%%%%%%%%%%%%%%%%%%%%%%%%%%%%%

%%%%%%%%%%%%%%%%%%%%%%%%%%%%%%%%%%%%%%%%%%%%%%%%%%%%%%%%%%%%%%%%%%%%%%%%%%%%
%%FIGURE 9 - 2 PI - SNAPSHOTS OF BLACK AND WHITE BOUNDARY MOVIE (1/2)
%%%%%%%%%%%%%%%%%%%%%%%%%%%%%%%%%%%%%%%%%%%%%%%%%%%%%%%%%%%%%%%%%%%%%%%%%%%%

\begin{figure*}[!htbp]

\centering\includegraphics[width=\textwidth]{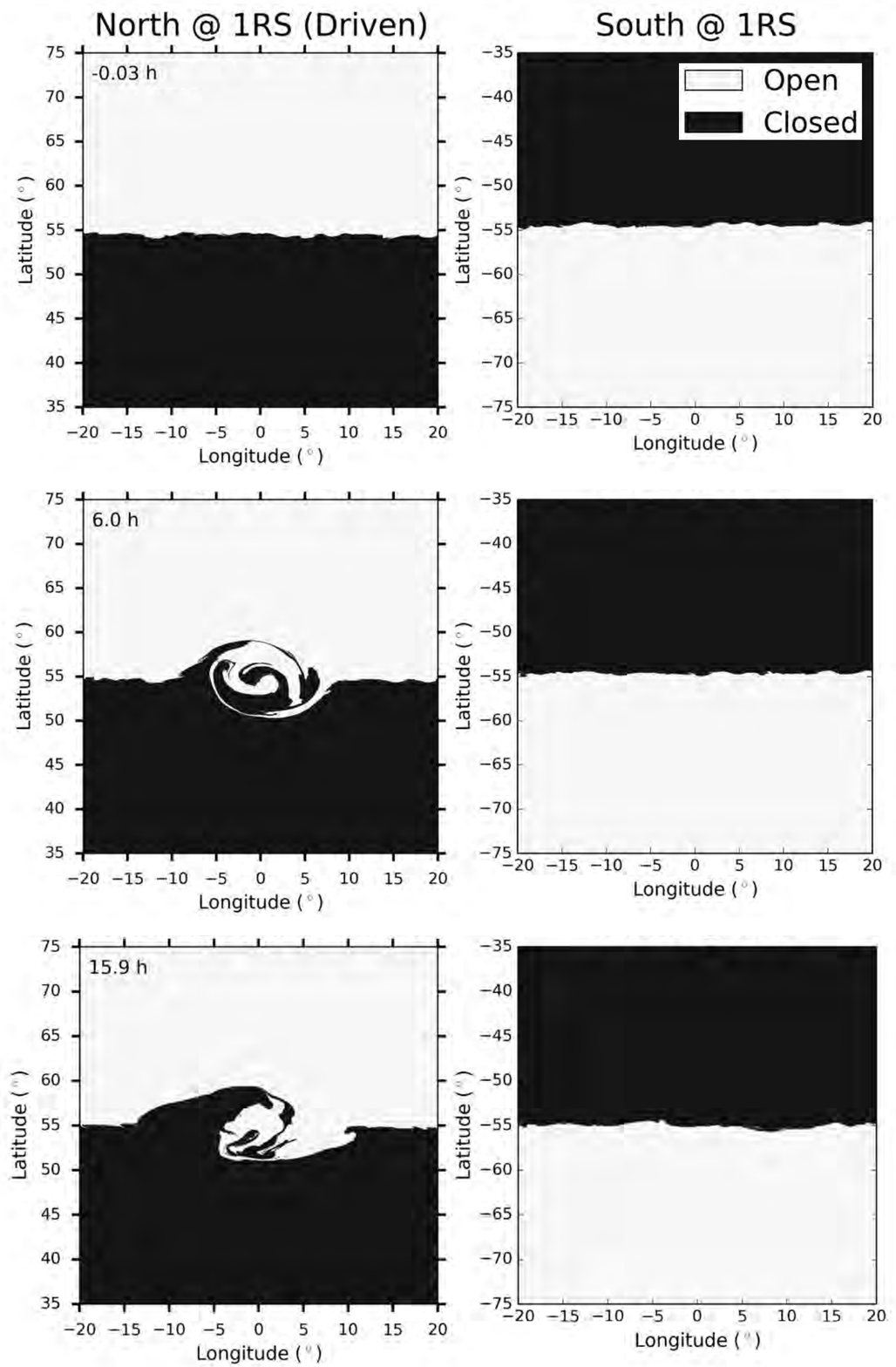}

\caption{Coronal-hole boundary maps for a peak displacement of $2\pi$, in the north 
(left) and south (right). White is open magnetic field, and black is closed. Top: 
Before driving, at $t = -0.03$ \hrs. Middle: End of driving, at $t = 6.0$ \hrs. 
Bottom: At $t = 15.9$ \hrs. The full, 5-minute-cadence movie is available online.}

\label{figure9}
%%%%%%%%%%%%%%%%%%%%%%%%%%%%%%%%%%%%%%%%%%%%%%%%%%%%%%%%%%%%%%%%%%%%%%%%%%%%
\end{figure*}
%%%%%%%%%%%%%%%%%%%%%%%%%%%%%%%%%%%%%%%%%%%%%%%%%%%%%%%%%%%%%%%%%%%%%%%%%%%%

%%%%%%%%%%%%%%%%%%%%%%%%%%%%%%%%%%%%%%%%%%%%%%%%%%%%%%%%%%%%%%%%%%%%%%%%%%%%
%%FIGURE 10 - 2 PI - SNAPSHOTS OF BLACK AND WHITE BOUNDARY MOVIE (2/2)
%%%%%%%%%%%%%%%%%%%%%%%%%%%%%%%%%%%%%%%%%%%%%%%%%%%%%%%%%%%%%%%%%%%%%%%%%%%%

\begin{figure*}[!htbp]

\centering\includegraphics[width=\textwidth]{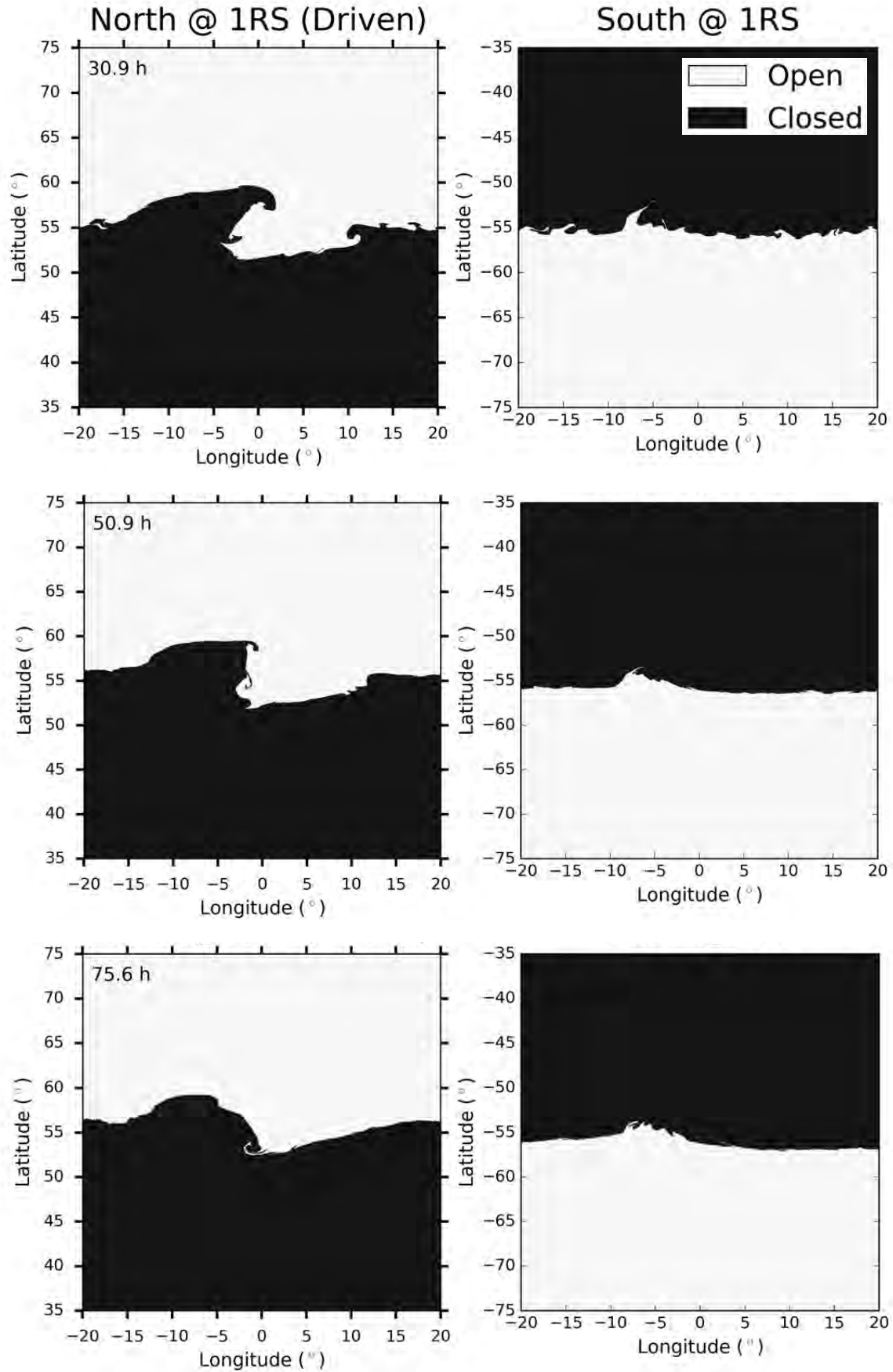}

\caption{Continuation of Figure \ref{figure9}. Top: At $t = 30.9$ \hrs. Middle: At 
$t=50.9$ \hrs. Bottom: At $t=75.6$ \hrs. The full, 5-minute-cadence movie is available 
online.}

\label{figure10}
%%%%%%%%%%%%%%%%%%%%%%%%%%%%%%%%%%%%%%%%%%%%%%%%%%%%%%%%%%%%%%%%%%%%%%%%%%%%
\end{figure*}
%%%%%%%%%%%%%%%%%%%%%%%%%%%%%%%%%%%%%%%%%%%%%%%%%%%%%%%%%%%%%%%%%%%%%%%%%%%%

%%%%%%%%%%%%%%%%%%%%%%%%%%%%%%%%%%%%%%%%%%%%%%%%%%%%%%%%%%%%%%%%%%%%%%%%%%%%
%%FIGURE 11 - 2 PI - LINE PLOT OF INTERCHANGE VS CLOSING FLUX 
%%%%%%%%%%%%%%%%%%%%%%%%%%%%%%%%%%%%%%%%%%%%%%%%%%%%%%%%%%%%%%%%%%%%%%%%%%%%

\begin{figure*}[!htbp]

\centering\includegraphics[width=\textwidth]{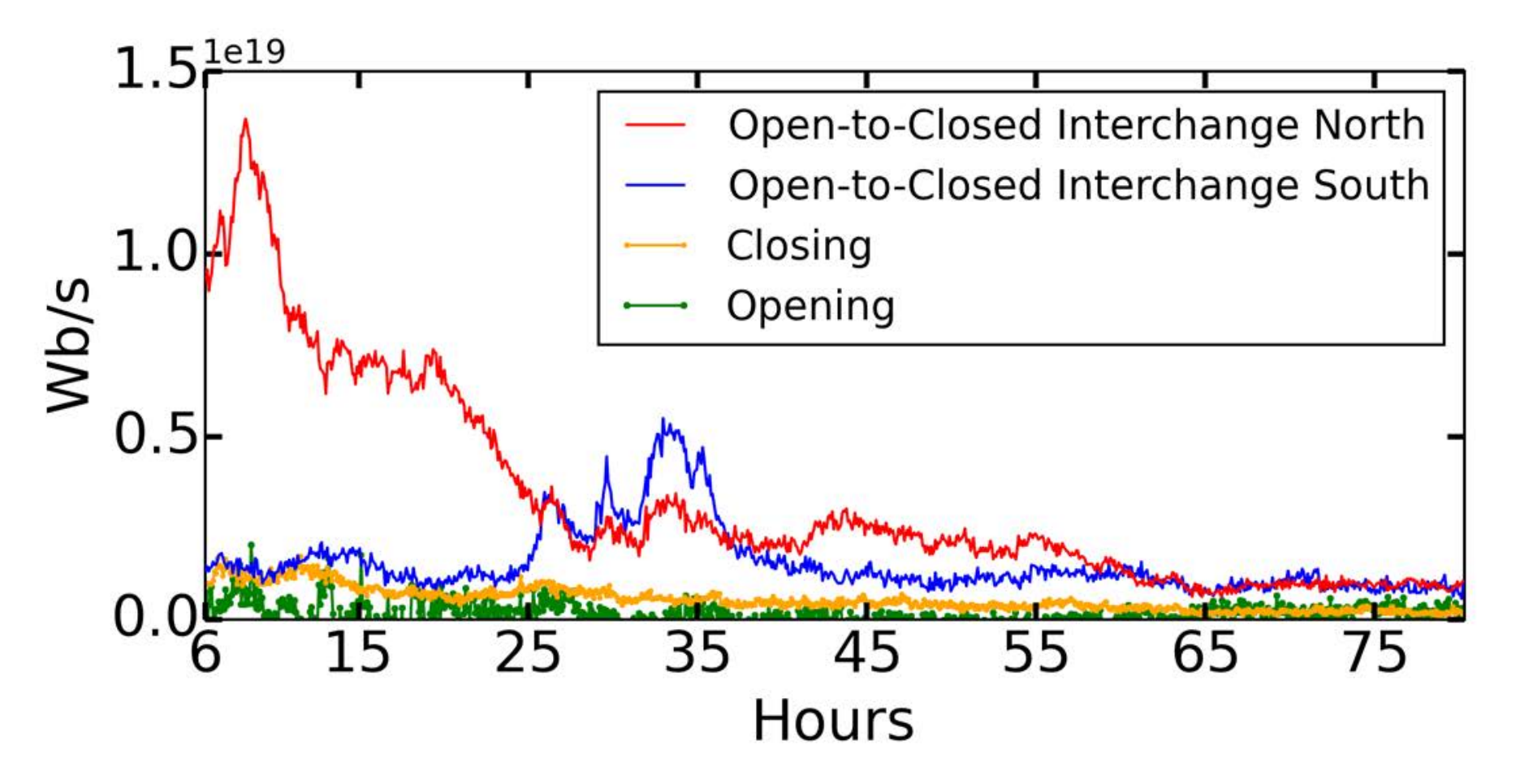}

\caption{Rate of flux change for a peak displacement of $2\pi$, where time $t = 6.0$ \hrs \;
corresponds to the end of the driving. Red and blue curves display the rate of flux change 
due to interchange reconnection in the north and south, respectively; yellow and green curves 
display the rate of flux change due to closing and opening, respectively. The rates of opening 
and closing are identical in the north and south.}

\label{figure11}
%%%%%%%%%%%%%%%%%%%%%%%%%%%%%%%%%%%%%%%%%%%%%%%%%%%%%%%%%%%%%%%%%%%%%%%%%%%%
\end{figure*}
%%%%%%%%%%%%%%%%%%%%%%%%%%%%%%%%%%%%%%%%%%%%%%%%%%%%%%%%%%%%%%%%%%%%%%%%%%%%

%%%%%%%%%%%%%%%%%%%%%%%%%%%%%%%%%%%%%%%%%%%%%%%%%%%%%%%%%%%%%%%%%%%%%%%%%%%%
%%FIGURE 12 - 2 PI - LAST FRAME OF ACCUMULATED INTERCHANGES PLOT 
%%%%%%%%%%%%%%%%%%%%%%%%%%%%%%%%%%%%%%%%%%%%%%%%%%%%%%%%%%%%%%%%%%%%%%%%%%%%

\begin{figure*}[!htbp]

\centering\includegraphics[width=\textwidth]{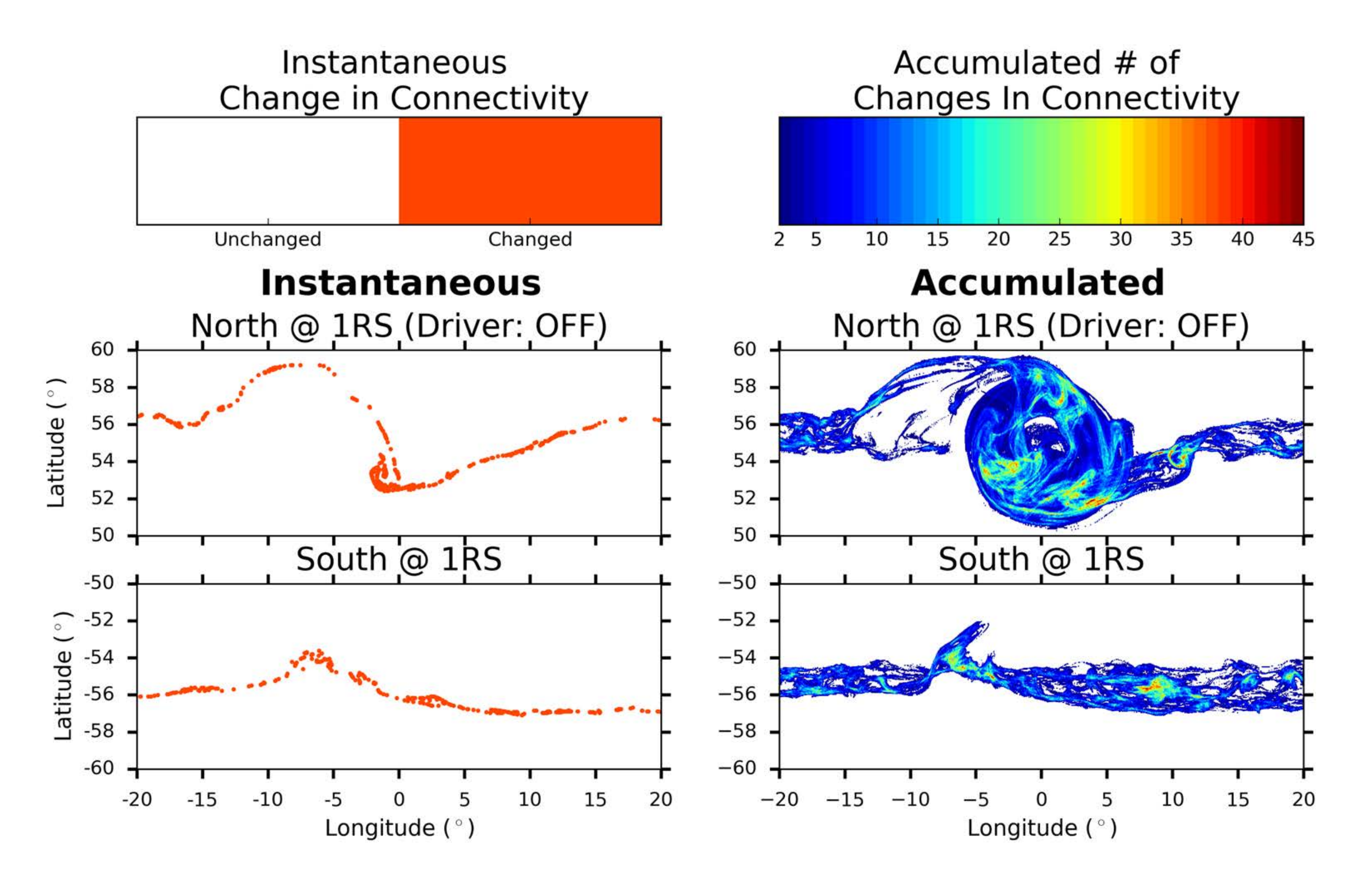}

\caption{Left: Instantaneous change in connectivity for a peak displacement of $2\pi$, 
in the north (top) and south (bottom). Red circles indicate the locations of field lines 
that have changed their connectivity between the previous and current snapshot. Shown here 
are the final maps at $t = 75.6$ \hrs. Right: Contour plot of the number of times each field 
line has changed connectivity over the entire duration of the simulation, in the north (top) 
and south (bottom). Field lines that have changed connectivity only once are not shown. 
The full, 5-minute-cadence movie is available online.}

\label{figure12}
%%%%%%%%%%%%%%%%%%%%%%%%%%%%%%%%%%%%%%%%%%%%%%%%%%%%%%%%%%%%%%%%%%%%%%%%%%%%
\end{figure*}
%%%%%%%%%%%%%%%%%%%%%%%%%%%%%%%%%%%%%%%%%%%%%%%%%%%%%%%%%%%%%%%%%%%%%%%%%%%%

%%%%%%%%%%%%%%%%%%%%%%%%%%%%%%%%%%%%%%%%%%%%%%%%%%%%%%%%%%%%%%%%%%%%%%%%%%%%
%%FIGURE 13 - 2 PI - SNAPSHOTS OF FOUR COLOR DOT PLOT (1/2) 
%%%%%%%%%%%%%%%%%%%%%%%%%%%%%%%%%%%%%%%%%%%%%%%%%%%%%%%%%%%%%%%%%%%%%%%%%%%%

\begin{figure*}[!htbp]

\centering\includegraphics[width=\textwidth]{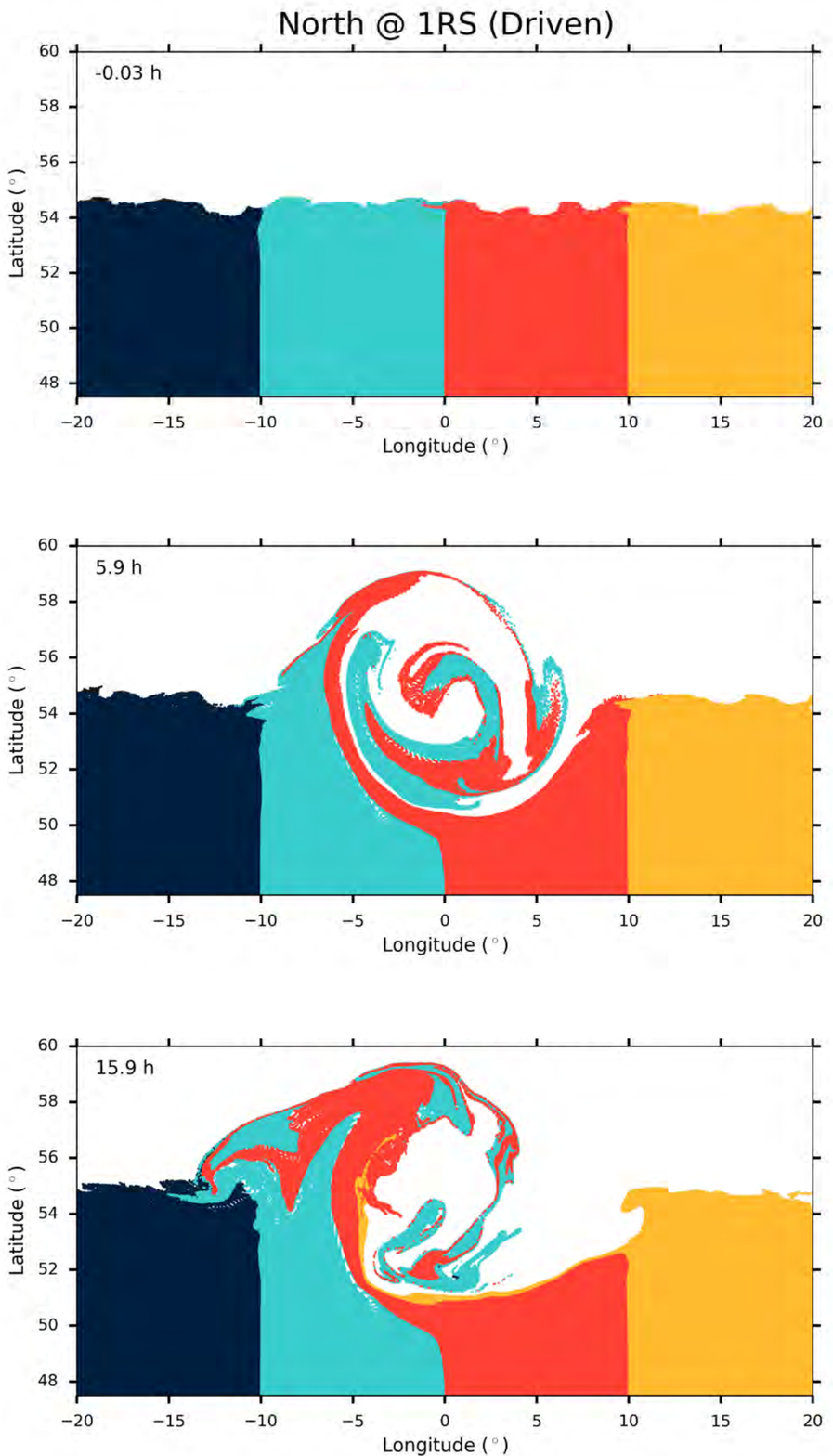}

\caption{Closed field lines in the north for a peak displacement of $2\pi$. Field lines 
are color-shaded based on the longitudes of their conjugate footpoints in the southern 
hemisphere: navy blue between $-20^\circ$ and $-10^\circ$; teal between $-10^\circ$ and 
$0^\circ$; red between $0^\circ$ and $+10^\circ$; and yellow between $+10^\circ$ and 
$+20^\circ$. Top: Before driving at $t = -0.03$ \hrs. Middle: End of driving at $t = 
6.0$ \hrs. Bottom: At $t = 15.9$ \hrs. The full, 5-minute-cadence movie is available 
online.}

\label{figure13}
%%%%%%%%%%%%%%%%%%%%%%%%%%%%%%%%%%%%%%%%%%%%%%%%%%%%%%%%%%%%%%%%%%%%%%%%%%%%
\end{figure*}
%%%%%%%%%%%%%%%%%%%%%%%%%%%%%%%%%%%%%%%%%%%%%%%%%%%%%%%%%%%%%%%%%%%%%%%%%%%%

%%%%%%%%%%%%%%%%%%%%%%%%%%%%%%%%%%%%%%%%%%%%%%%%%%%%%%%%%%%%%%%%%%%%%%%%%%%%
%%FIGURE 14 - 2 PI - SNAPSHOTS OF FOUR COLOR DOT PLOT (2/2) 
%%%%%%%%%%%%%%%%%%%%%%%%%%%%%%%%%%%%%%%%%%%%%%%%%%%%%%%%%%%%%%%%%%%%%%%%%%%%

\begin{figure*}[!htbp]

\centering\includegraphics[width=\textwidth]{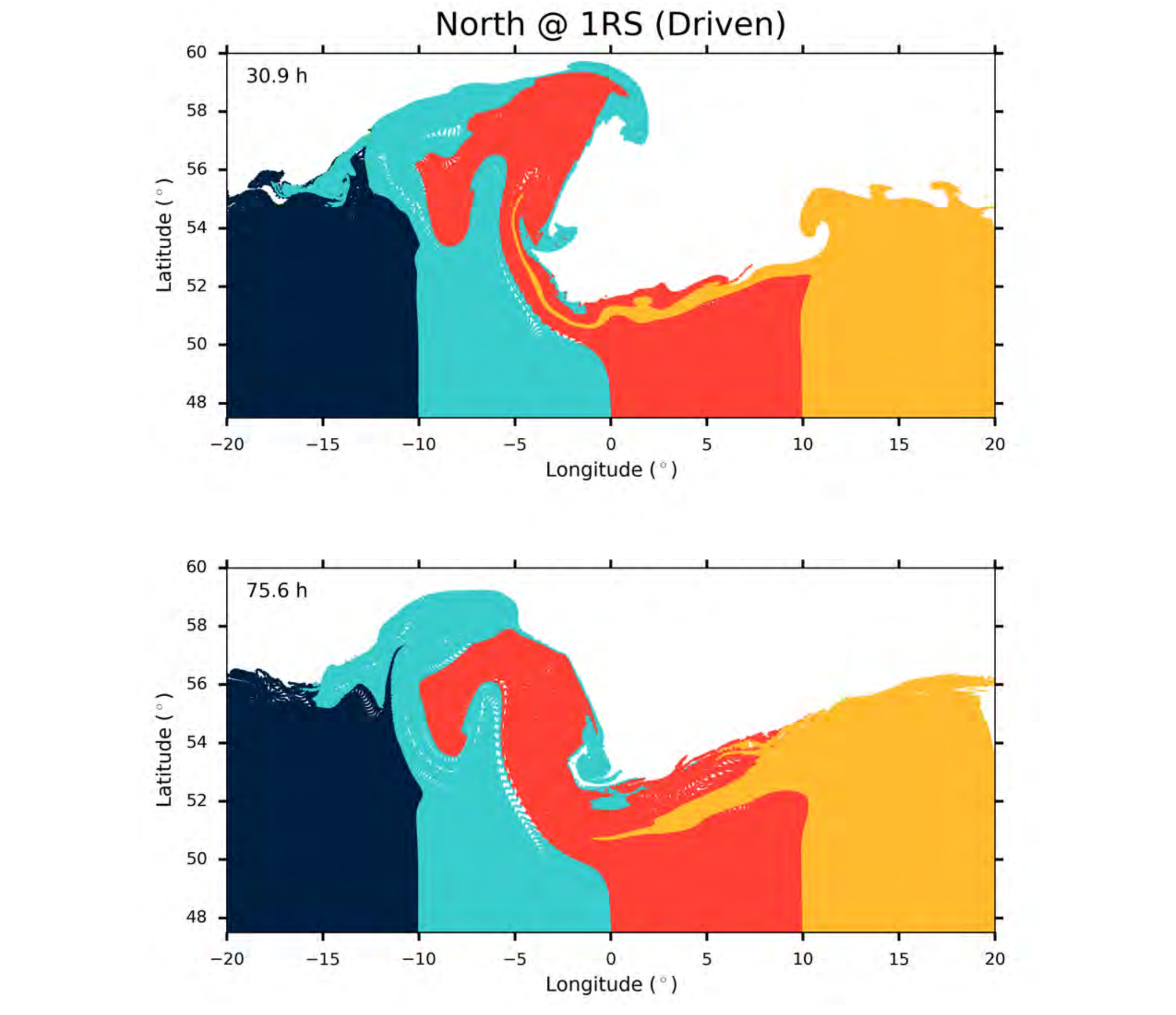}

\caption{Continuation of Figure \ref{figure13}, using the same color scheme. Top: At 
$t = 30.9$ \hrs. Bottom: At $t = 75.6$ \hrs. The full, 5-minute-cadence movie is 
available online.}

\label{figure14}
%%%%%%%%%%%%%%%%%%%%%%%%%%%%%%%%%%%%%%%%%%%%%%%%%%%%%%%%%%%%%%%%%%%%%%%%%%%%
\end{figure*}
%%%%%%%%%%%%%%%%%%%%%%%%%%%%%%%%%%%%%%%%%%%%%%%%%%%%%%%%%%%%%%%%%%%%%%%%%%%%

%%%%%%%%%%%%%%%%%%%%%%%%%%%%%%%%%%%%%%%%%%%%%%%%%%%%%%%%%%%%%%%%%%%%%%%%%%%%
%%FIGURE 15 - CARTOON
%%%%%%%%%%%%%%%%%%%%%%%%%%%%%%%%%%%%%%%%%%%%%%%%%%%%%%%%%%%%%%%%%%%%%%%%%%%%

\begin{figure*}[!htbp]

\centering\includegraphics[width=\textwidth]{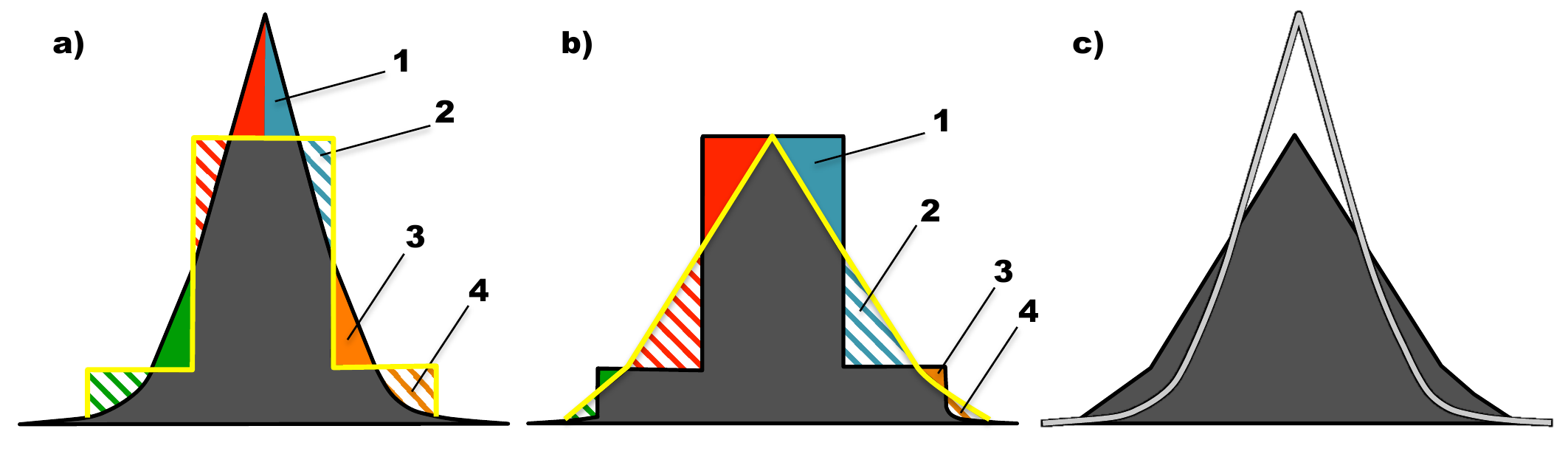}

\caption{Cartoon schematic of interchange reconnection changing the boundary of 
a closed-field region. Thick black lines: old coronal-hole 
boundary. Thick yellow line: new coronal-hole boundary. Thick gray line: original 
corona-hole boundary. Solid-shaded gray, red, blue, green and orange areas 
represent closed-field regions. Striped red, blue, green, and orange areas 
represent open-field regions. Panels a, b, and c illustrate the evolution 
over time as closed regions 1 and 3 experience interchange reconnection with the open regions 2 and 4, respectively. See text for details. }

\label{figure15}
%%%%%%%%%%%%%%%%%%%%%%%%%%%%%%%%%%%%%%%%%%%%%%%%%%%%%%%%%%%%%%%%%%%%%%%%%%%%
\end{figure*}
%%%%%%%%%%%%%%%%%%%%%%%%%%%%%%%%%%%%%%%%%%%%%%%%%%%%%%%%%%%%%%%%%%%%%%%%%%%%

%%%%%%%%%%%%%%%%%%%%%%%%%%%%%%%%%%%%%%%%%%%%%%%%%%%%%%%%%%%%%%%%%%%%%%%%%%%%
%%FIGURE 16 - 2 PI - EXAMPLE OF AN INTERCHANGE EVENTS
%%%%%%%%%%%%%%%%%%%%%%%%%%%%%%%%%%%%%%%%%%%%%%%%%%%%%%%%%%%%%%%%%%%%%%%%%%%%

\begin{figure}[!htbp]

\includegraphics[width=0.5\textwidth]{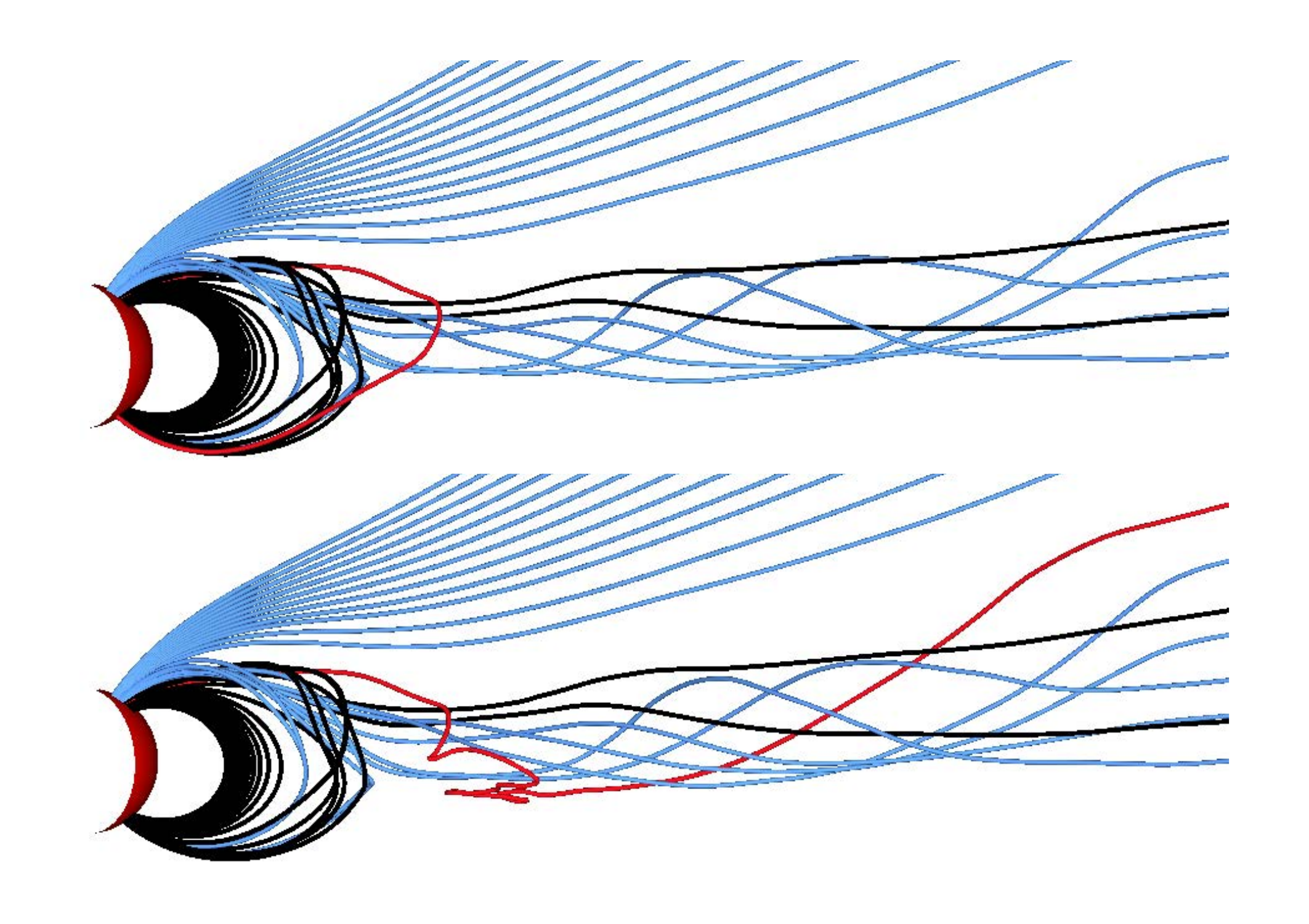}

\caption{Example of an individual interchange event from the $2\pi$ displacement case. 
Blue field lines were originally open; black field lines were originally closed; red 
field line interchange-reconnects from closed to open between the top and bottom panels, 
which are 5 minutes apart. See text for details.}

\label{figure16}
%%%%%%%%%%%%%%%%%%%%%%%%%%%%%%%%%%%%%%%%%%%%%%%%%%%%%%%%%%%%%%%%%%%%%%%%%%%%
\end{figure}
%%%%%%%%%%%%%%%%%%%%%%%%%%%%%%%%%%%%%%%%%%%%%%%%%%%%%%%%%%%%%%%%%%%%%%%%%%%%

%%%%%%%%%%%%%%%%%%%%%%%%%%%%%%%%%%%%%%%%%%%%%%%%%%%%%%%%%%%%%%%%%%%%%%%%%%%%
%%FIGURE 17 - 2 PI - WIDTH OF DYNAMIC REGION BEFORE AND AFTER FLOW  
%%%%%%%%%%%%%%%%%%%%%%%%%%%%%%%%%%%%%%%%%%%%%%%%%%%%%%%%%%%%%%%%%%%%%%%%%%%%

\begin{figure*}[!htbp]

\centering\includegraphics[width=\textwidth]{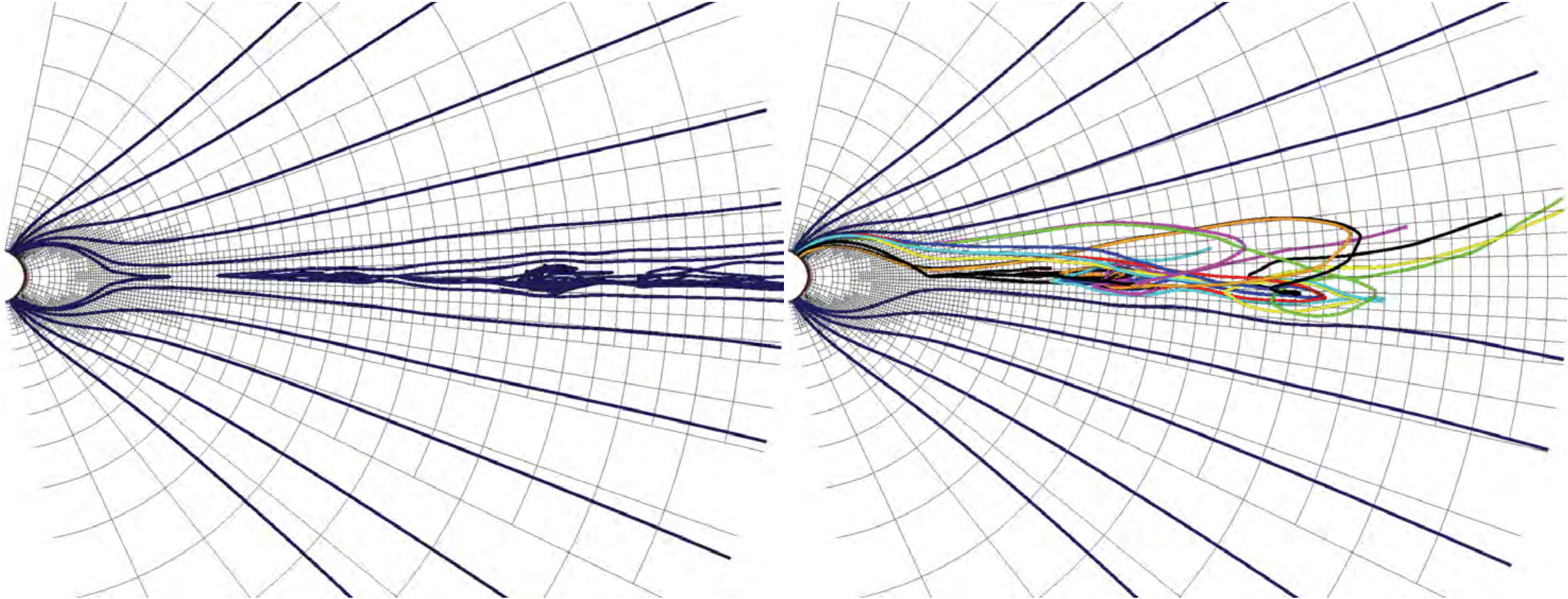}

\caption{Change in width of the dynamic slow-wind region for the $2\pi$ displacement 
case. Left: Blue field lines in dynamic equilibrium show a narrow, $\sim 5^\circ$ 
width of the slow solar wind. Right: Multi-colored field lines extend to much higher 
latitudes due to interchange reconnection driven by the photospheric flow, showing 
a broad, $\sim 15^\circ$ width of the slow solar wind.}

\label{figure17}
%%%%%%%%%%%%%%%%%%%%%%%%%%%%%%%%%%%%%%%%%%%%%%%%%%%%%%%%%%%%%%%%%%%%%%%%%%%%
\end{figure*}
%%%%%%%%%%%%%%%%%%%%%%%%%%%%%%%%%%%%%%%%%%%%%%%%%%%%%%%%%%%%%%%%%%%%%%%%%%%%

%%%%%%%%%%%%%%%%%%%%%%%%%%%%%%%%%%%%%%%%%%%%%%%%%%%%%%%%%%%%%%%%%%%%%%%%%%%%
\end{document}